\documentclass[%
 aps,
 prb,
 reprint,
 amsmath,amssymb,
]{revtex4-2}

\usepackage[export]{adjustbox} 
\usepackage{graphicx}
\usepackage{dcolumn}
\usepackage{bm}

\usepackage{graphicx}
\graphicspath{ {./Figures/} }
\usepackage{xfrac}
\usepackage{amsmath}
\usepackage{amssymb}
\usepackage{amsfonts}
\usepackage{caption}
\usepackage{diagbox} 
\usepackage{subcaption}
\usepackage{float}
\usepackage{color}
\usepackage{mathtools}
\usepackage{hyperref}
\usepackage{url}
\usepackage{bm}
\usepackage{eqnarray}

\usepackage{amsthm}

\usepackage[table,xcdraw]{xcolor}

\usepackage[makeroom]{cancel} 

\hypersetup{
  colorlinks   = true, 
  urlcolor     = blue, 
  linkcolor    = blue, 
  citecolor    = red 
}
\usepackage{braket}

\usepackage[table,xcdraw]{xcolor}
\usepackage{color, colortbl}

\usepackage{relsize}

\begin{document}

\setlength{\abovedisplayskip}{5pt}
\setlength{\belowdisplayskip}{5pt}

\preprint{APS/123-QED}

\title{Strain-tunable magnetic correlations in spin liquid candidate Nb$_3$Cl$_8$}

\author{Tharindu Fernando}
 \email{tharindu@uw.edu}
 \affiliation{%
 Department of Physics, University of Washington, Seattle, WA 98195 USA
}%
\author{Ting Cao}%
 \affiliation{Department of Materials Science and Engineering, University of Washington, Seattle, WA 98195 USA}

\date{\today}

\begin{abstract}
Recent research suggests the possibility of the two-dimensional breathing-Kagome magnet Nb$_3$Cl$_8$ 
hosting a quantum spin liquid state, warranting further study into its magnetic properties. 
Using \emph{ab initio} calculations, we show that monolayer Nb$_3$Cl$_8$ has short-range antiferromagnetic correlations among Nb$_3$ trimers with S = 1/2, and becomes  
magnetically frustrated due to the underlying effective triangular lattice geometry, and is evidenced by a
frustration index of $f>1$. 
The high-temperature susceptibility shows a negative Weiss temperature from Monte Carlo calculations. 
Considering spin-orbit coupling, we investigate the magnetic anisotropy, including anisotropic exchange, single-ion anisotropy and the 
Dzyaloshinskii–Moriya interaction using the four-state energy mapping formalism.
Although the elements have relatively small atomic numbers, the Dzyaloshinskii–Moriya interaction is comparable in magnitude to the anisotropic exchange.
Additionally, 
we show that biaxial strain tunes 
the short-range correlations
between antiferromagnetic, paramagnetic and ferromagnetic.
These findings strengthen our understanding of Nb$_3$Cl$_8$
and advance its applications in current condensed matter physics and materials science research, including nanoscale mechanical and spintronics applications.
\end{abstract}

         
\maketitle

\newpage

\section{Introduction}
The study of two-dimensional (2D) magnetic phenomena in quantum materials has attracted significant interest due to its potential to reveal novel quantum states and exotic magnetic properties \cite{burch2018magnetism,jiang2021recent,zhang2021two}. 
A central goal is to control and tune magnetic order, which not only enables the exploration of new physical phenomena but also drives the development of advanced technologies and the improvement of existing devices \cite{li2021tuning,yu2022recent}. 
Magnetic tunability underpins applications in condensed matter physics and materials science, including quantum information science and technology \cite{vzutic2004spintronics,zhang20242d}, spintronic devices and memory \cite{vzutic2004spintronics,zhang20242d}, magnetic sensing and imaging \cite{degen2017quantum}, and the study of exotic electronic and magnetic phases \cite{burch2018magnetism,jiang2021recent,zhang2021two}. 
Frustrated magnets, where competing interactions prevent the system from adopting a simple order, are especially intriguing. They can host exotic states such as spin liquids, which provide insights into high-temperature superconductivity, quantum computation, and other emergent phenomena \cite{mendels2011introduction,balents2010spin,broholm2020quantum}.  

\begin{figure*}
    \centering
    \includegraphics[width=0.6\textwidth]{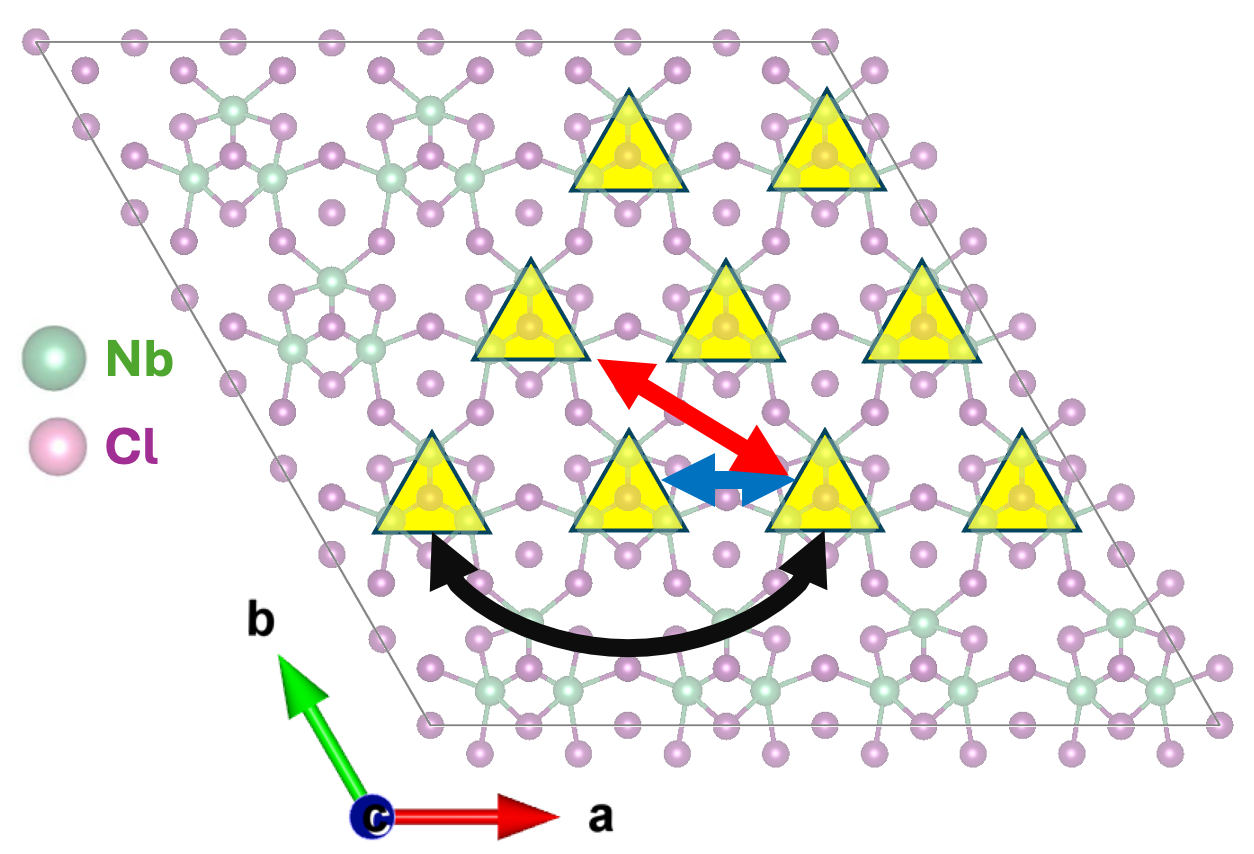}
    \caption{
        Nb$_3$Cl$_8$ crystal structure shown as a $4\times 4\times 1$ supercell, viewed along the $c$-axis \cite{momma2008vesta}.
        Semi-transparent spheres represent Nb (green) and Cl (purple) atoms of the breathing Kagome lattice, with alternating large and small Nb$_3$ triangles. 
        Selected Nb$_3$ trimers (small triangles, each hosting a single $S=1/2$ moment) are highlighted in yellow.
        Consider each yellow trimer as a single unit that produces the overlying triangular lattice. 
        Arrows indicate NN (blue), 2NN (red), and 3NN (black) trimer-trimer interactions used in our calculations. 
        The triangle with three arrows pointing to it is taken as the first index of each bond.
    }
    \label{fig:lattice}
\end{figure*}

Niobium halides Nb$_3$X$_8$ (X = Cl, Br, I), which are van der Waals (vdW) materials, are promising candidates for exploring such quantum phenomena. 
Their breathing Kagome lattice comprises alternating large and small Nb$_3$ triangles. 
Each small triangle, or \emph{Nb$_3$ trimer}, hosts a shared molecular orbital with a $S=1/2$ moment \cite{sheckelton2015strongly,haraguchi2017magnetic,pasco2019tunable,sheckelton2017rearrangement,hu2023correlated,liu2024possible,grytsiuk2024nb3cl8,gao2023discovery,zhang2023mottness,stepanov2024signatures,yoon2020anomalous,haraguchi2024molecular,nakamura2024charge}. 
Thus, despite the breathing Kagome geometry, Nb$_3$X$_8$ effectively behaves as a triangular lattice magnet, with each trimer corresponding to one triangle's vertex (Fig.~\ref{fig:lattice}). 
The triangular lattice geometry naturally supports magnetic frustration, since spins on an equilateral triangle cannot all align antiferromagnetically. 
If its magnetic correlations could be tuned between antiferromagnetic and other states, Nb$_3$X$_8$ would provide an appealing platform for realizing controllable quantum magnetism. 
Accurate modeling of its magnetic properties is therefore essential for assessing its potential to host exotic quantum states.  

In this work, we focus on Nb$_3$Cl$_8$, since our \emph{ab initio} screening showed an antiferromagnetic nearest-neighbor exchange ($J^1$), in contrast to the ferromagnetic $J^1$ found in Nb$_3$Br$_8$ and Nb$_3$I$_8$ (see also Ref. \cite{mangeri2024magnetoelectric}). The presence of an antiferromagnetic $J^1$ on an effective triangular lattice geometry makes Nb$_3$Cl$_8$ a particularly interesting case for detailed study, as it introduces competing interactions that can lead to enhanced magnetic frustration and unconventional magnetic ground states.

Several additional factors motivate our choice of Nb$_3$Cl$_8$. 
Nb$_3$Cl$_8$ may host topological flat bands in its electronic dispersion \cite{liu2021screening,regnault2022catalogue,cualuguaru2022general,duan2022inventory,jeff2023raman,sun2022observation}, which in turn can stabilize strongly correlated insulating phases \cite{hu2023correlated,sheckelton2015strongly,aretz2025strong}. 
It has also been identified as a candidate to realize the Hubbard model \cite{zhang2023mottness,grytsiuk2024nb3cl8,gao2023discovery}, Mott-insulating behavior \cite{hu2023correlated,liu2024possible,grytsiuk2024nb3cl8,gao2023discovery,zhang2023mottness,stepanov2024signatures,yang2025gate}, field-effect transistors \cite{lu2024bipolar,lee2024negative}, topological insulators \cite{bradlyn2017topological}, heterostructure phenomena \cite{lu2025intrinsic,yang2025electronic}, excitonic effects \cite{khan2024multiferroic}, and charge-ice physics \cite{stepanov2024signatures}. 
Magnetically, both bulk and monolayer Nb$_3$Cl$_8$ are paramagnetic at high temperature \cite{gao2023discovery,pasco2019tunable,stepanov2024signatures}, with the bulk undergoing a $\sim\!90$~K transition to a nonmagnetic singlet ground state \cite{haraguchi2017magnetic,sheckelton2017rearrangement,pasco2019tunable,kim2023terahertz,gao2023discovery,jeff2023raman,bouhmouche2024highly,liu2024possible,haraguchi2024molecular}. 
However, the exact magnetic ground state of the monolayer remains unsettled, with reports ranging from ferromagnetic \cite{jiang2017exploration,kim2023terahertz} to antiferromagnetic \cite{sun2022observation,stepanov2024signatures,haraguchi2024molecular,yoon2020anomalous,grytsiuk2024nb3cl8,aretz2025strong} and even quantum spin liquid candidates \cite{hu2023correlated,liu2024possible,haraguchi2024molecular,grytsiuk2024nb3cl8,zhang2023mottness}. 
To date, there are very few studies addressing the impact of spin–orbit coupling (SOC) in the monolayer or exploring how its magnetic correlations can be tuned. 
Strain is especially important, as it can drive changes in the magnetic ground state as shown in CrSBr \cite{cenker2022reversible}; yet aside from Ref.~\cite{mortazavi2022first}, which considered strain only in terms of mechanical properties, strain effects in Nb$_3$Cl$_8$ have not been examined.  

In this work, we explore the magnetic properties of monolayer Nb$_3$Cl$_8$. 
Recent studies of the monolayer cover a wide range of topics \cite{hu2023correlated,grytsiuk2024nb3cl8,gao2023discovery,mortazavi2022first,sun2022observation,zhang2023mottness,stepanov2024signatures,jiang2017exploration,yoon2020anomalous}, including evidence that its synthesis may be experimentally achievable \cite{sun2022observation,mortazavi2022first,jiang2017exploration,yoon2020anomalous}. 
Nb$_3$Cl$_8$ belongs to space group $156$ (P3m1), which lacks inversion symmetry, permitting nonzero Dzyaloshinskii–Moriya interactions (DMI) in the presence of SOC \footnote{Band structure and magnetism diagrams are provided in Supplementary Material I.}. 
SOC further enriches the magnetic landscape by introducing anisotropy and extending the role of extended-neighbor interactions, whose combined influence on magnetic correlations remains unexplored.  

A common approach to study magnetic anisotropy is to calculate energy differences among distinct spin configurations. 
The four-state energy mapping formalism is especially powerful in this regard \cite{xiang2011predicting,xiang2013magnetic,vsabani2020ab}. 
It reduces ambiguity in near-degenerate cases, accounts comprehensively for interactions (including anisotropic exchange), and, when based on accurate \emph{ab initio} calculations, can yield good agreement with experiment. 
Yet for Nb$_3$Cl$_8$, this formalism has not been applied to the effective triangular lattice formed by Nb$_3$ trimers \cite{zhang2023topological,sun2022observation,grytsiuk2024nb3cl8,haraguchi2017magnetic,gao2023discovery,sheckelton2017rearrangement,stepanov2024signatures,aretz2025strong,mangeri2024magnetoelectric}.  

To address this gap, we employ the four-state mapping method \cite{vsabani2020ab} on the triangular trimer lattice to extract magnetic anisotropy parameters for a generalized Heisenberg-like Hamiltonian [Eq.~\eqref{eq:hamiltonian}], including interactions up to third-nearest neighbors. 
To capture the system’s complexity more accurately, we extend beyond isotropic exchange \cite{grytsiuk2024nb3cl8,haraguchi2017magnetic,gao2023discovery,sheckelton2017rearrangement,stepanov2024signatures,zhang2023topological,sun2022observation,aretz2025strong,mangeri2024magnetoelectric} by incorporating SOC-induced anisotropic exchange (including DMI), and single-ion anisotropy. 
Using \emph{ab initio} calculations, we determine these parameters within a unified and internally consistent framework, and subsequently calculate the magnetic susceptibility using classical Monte Carlo simulations. 
Fitting the data to the Curie-Weiss law reveals short-range antiferromagnetic correlations with a negative Weiss temperature, consistent with frustration on the triangular lattice (that is also evidenced by a frustration index of $f>1$). 
We also find finite DMI of comparable magnitude to the anisotropic exchange. 
Finally, we demonstrate that biaxial strain can tune the magnetism between antiferromagnetic, paramagnetic, and ferromagnetic local correlations, providing a route to manipulate the magnetic properties of Nb$_3$Cl$_8$.

\section{Methods}
With the anisotropies allowed by SOC, and extended-neighbor interactions, 
the corresponding generalized Heisenberg-like spin Hamiltonian becomes:
\begin{equation}
\begin{aligned}\label{eq:hamiltonian}
H = 
&\sum_{i<j}  \mathbf{S}_i \cdot J^k_{ij}\cdot \mathbf{S}_j 
+\sum_{i} A_i (S_i^z)^2
\end{aligned}
\end{equation}
where
$k=1,2,3$ corresponds to 
nearest-neighbor (NN),  
second-nearest-neighbor (2NN) 
and third-nearest-neighbor (3NN) pairs respectively. 
We use the index $k$ to make explicit the usual $J^1$, $J^2$, $J^3$ notation for NN–3NN couplings.
\(\mathbf{S}_i = (S_i^x, S_i^y, S_i^z)\) represents the spin-1/2 at site \(i\).
\(J^k_{ij}\) is the anisotropic exchange interaction between spins at sites \(i\) and \(j\), and
\(A_i\) is the single-ion anisotropy constant at site \(i\). This form of the 
single-ion anisotropy contribution to the 
Hamiltonian is due to the system's easy-plane
anisotropy (see next section) \footnote{Otherwise, we would have the 
general form 
$A_{i,xx} (S_i^x)^2 + A_{i,yy} (S_i^y)^2 + A_{i,zz} (S_i^z)^2 + 2A_{i,xy} S_i^x S_i^y + 2A_{i,xz} S_i^x S_i^z + 2A_{i,yz} S_i^y S_i^z$ \cite{xiang2013magnetic}. 
In our case, we have $A_i (S^z)^2 \equiv (A_{i,zz}-A_{i,xx}) (S_i^z)^2$.}.  
From $J_{ij}^k$, we can get the components of the DMI
vectors $D^k=(D_x^k,D_y^k,D_z^k)$ in a Cartesian frame of reference
as $D_x^k = \tfrac12(J^k_{yz}-J^k_{zy}),\; D_y^k = \tfrac12(J^k_{zx}-J^k_{xz}),\; D_z^k = \tfrac12(J^k_{xy}-J^k_{yx})$.

We note that, to the best of our knowledge, the complete Hamiltonian in Eq. \eqref{eq:hamiltonian}, including anisotropic exchange, DMI, and single-ion anisotropy up to third-nearest neighbors, has not been systematically explored in previous studies of spin-$1/2$ triangular-lattice systems. Earlier works have typically considered simplified models, such as isotropic or XXZ-type exchanges, sometimes including DMI or up to only second-nearest-neighbor interactions, within limited parameter regimes \cite{saadatmand2015phase,glittum2021arc,bishop2015spin,gong2017global,jiang2023nature,van2024magnetic}. By using the full Hamiltonian, we can capture the combined influence of SOC-induced anisotropies and extended-neighbor couplings on the competition between ferromagnetic and antiferromagnetic correlations, the emergence of noncollinear spin textures, and possible chiral ground states, which cannot always be accessed within the simplified forms.

We used an \emph{ab initio} density functional theory (DFT) approach to perform first-principles DFT+U+SOC calculations. We leave details of our DFT calculations to Supplementary Material A.
We used the four-state energy mapping technique to calculate 
magnetic anisotropy parameters \cite{xiang2011predicting,xiang2013magnetic,vsabani2020ab}. 
In our calculations,
we used the trimer pairs indicated by arrows in
Fig. \ref{fig:lattice}
\footnote{The magnetic parameters between 
different trimer pairs will differ by rotations of the 
exchange matrices Eq. \eqref{eq:J-matrix-elems};
for instance, by $60^\circ$ for adjacent NN trimer pairs.}.

We used Monte Carlo calculations to obtain magnetic susceptibility as a function of temperature.
Then the Curie-Weiss law
identifies the role of dominant exchange interactions and suggest possible spin-spin correlations.
The magnetic susceptibility 
$\chi_d$ (in $d=x,y,z$ directions) of antiferromagnetic (AFM) and ferromagnetic (FM) 
materials above their respective critical temperatures (N\'eel temperature $T_N$ for antiferromagnets, and Curie temperature $T_C$ for ferromagnets) is:
\begin{equation}\label{eq:CW}
    \frac{1}{\chi_d} = \frac{T - \theta_d}{C}.
\end{equation} 
Above, we have the temperature $T$, Curie constant $C$ and Weiss temperature $\theta_d$.
$\theta_d$ is typically negative in AFM states,
0 in paramagnetic (PM) states,
and positive in FM states \cite{mugiraneza2022tutorial}.
Although anisotropy, as in the case of this work, may affect long-range magnetic order, the sign and magnitude of $\theta_d$ still reflect the presence of short-range correlations.
We refer the reader to Supplementary Material B for details on our Monte Carlo calculations.

\section{Frustration and Dzyaloshinskii–Moriya interaction}\label{sec:frustrationANDdmi}
For the triangle vertex pairs illustrated in Fig. \ref{fig:lattice}, 
we calculated the following (in units of meV):

\begin{equation}\label{eq:results_J123}
\begin{array}{c}
\begin{aligned}
J^1 = 
\begin{bmatrix}
  1.49 & -0.15 &  0.89 \\
  0.15 &  1.59 & -0.01 \\
 -0.89 & -0.01 &  1.47
\end{bmatrix}, 
J^2 = 
\begin{bmatrix}
  0.96 & 0.00 &  0.04 \\
  0.00 & 0.97 & -0.02 \\
 -0.04 & 0.02 &  0.95
\end{bmatrix}
\end{aligned}
\vspace{5pt}
\\[1ex]
\begin{aligned}
J^3 &= 
\begin{bmatrix}
 -0.73 & 0.00 & 0.00 \\
  0.00 & -0.74 & 0.00 \\
  0.00 & 0.00  & -0.73
\end{bmatrix}
\end{aligned}
\end{array}
\end{equation}
where the anisotropic exchange $J^k$ is
(using a Cartesian frame of reference, and 
ignoring the superscript $k=1,2,3$ for brevity):
\begin{equation}\label{eq:J-matrix-elems}
{J} = \begin{bmatrix}
J_{xx} & J_{xy} & J_{xz} \\
J_{yx} & J_{yy} & J_{yz} \\
J_{zx} & J_{zy} & J_{zz}
\end{bmatrix}.
\end{equation}
The single-ion anisotropy constant
was $A=0.56$ meV. 
The DMI vector $\mathbf{D}^k=(D_x,D_y,D_z)$
was calculated to be $\mathbf{D}^1=(0, -0.89, -0.15)$ meV,
$\mathbf{D}^2=(-0.02, -0.04, 0)$ meV,
and 
$\mathbf{D}^3=(0, 0, 0)$ meV (see Fig. S1).
Using the convention dictated by Eq. \eqref{eq:hamiltonian}, 
$J^k_{mn} > 0$ is an AFM interaction
(and $J^k_{mn} < 0$ is FM),
for $m,n$ being $x,y,z$.
$A > 0$ implies easy-plane anisotropy.

\begin{figure}[H]
    \centering
    \includegraphics[width=0.48\textwidth]{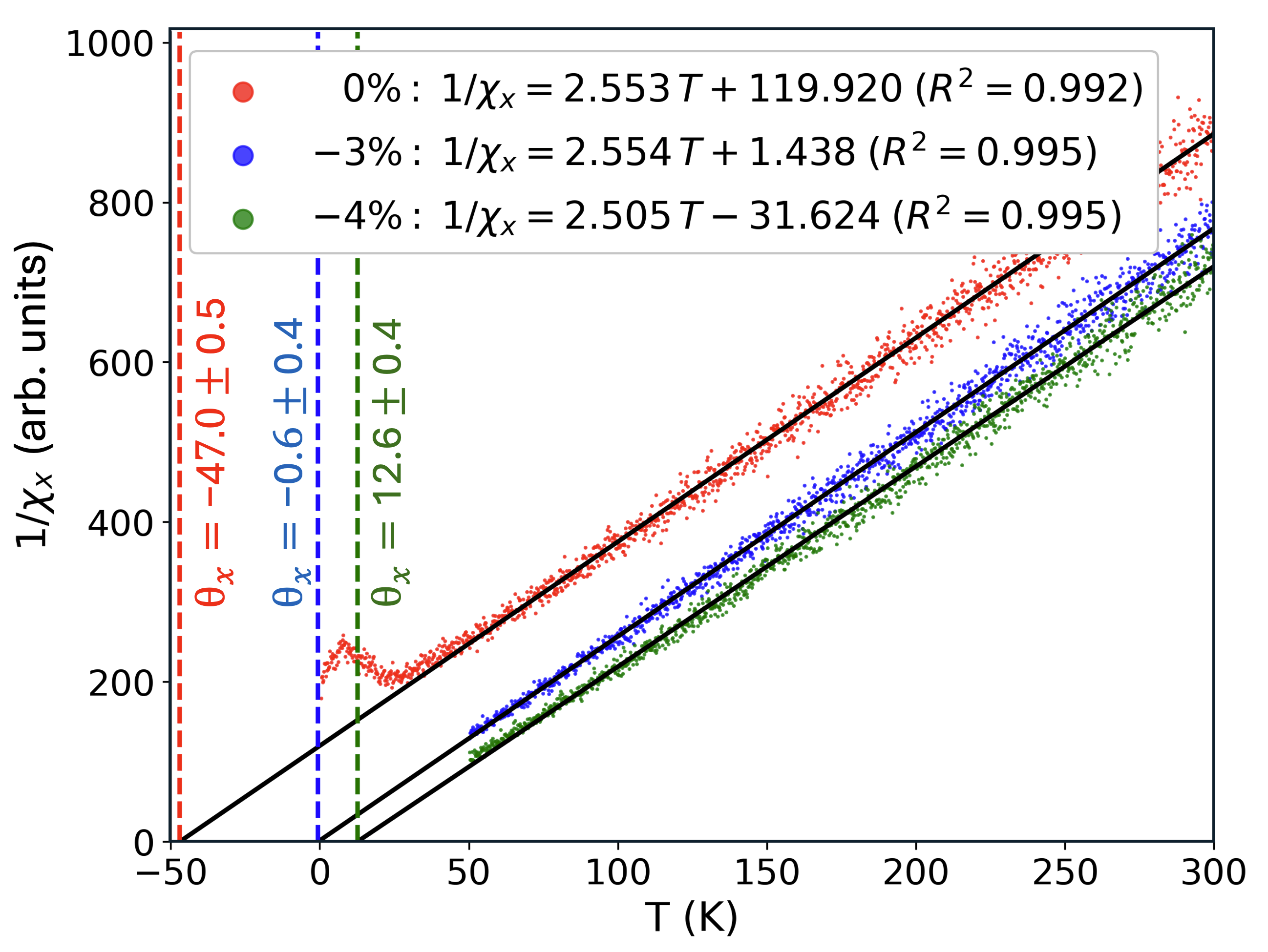}
    \caption{    
        Inverse magnetic susceptibility $1/\chi_d$ (arb. units) 
        vs. temperature $T (K)$ in the $d=x$-direction for
        $0\%$ (unstrained), $-3\%,$ and $-4\%$ biaxial strain.
        Data from Monte Carlo simulations are shown as dots. 
        For clarity in illustration, we 
        show Monte Carlo data for only the $50 \leq T \leq 300$ $K$ range 
        used in the fitting, with the 
        exception of $0\%$ strain, which exemplifies the kink indicating $T_N$.     
        The Weiss temperature $\theta_d$ and uncertainty
        for each case is given alongside the vertical dashed lines denoting the intersection of the linear fits with 
        the $T$ axis.    
    }
    \label{fig:sus_x}
\end{figure}

Notice that most of the nonzero components of 
$J^1$ and $J^2$ are $>0$, suggesting that
they are AFM. 
Indeed, using our classical Monte Carlo calculations at high-T,
we found the Weiss temperatures
$(\theta_x,\theta_y,\theta_z)
=(-47.0 \pm 0.5,$ $-46.9 \pm 0.5,$ $-48.8 \pm 0.5)$ 
$K$. $\theta_z$ is different from $\theta_x$ and $\theta_y$ due to the in-plane anisotropy.
The temperature dependent susceptibility used to calculate $\theta_d$ 
is given in Fig. \ref{fig:sus_x} and Fig. S4.
In all cases, we obtained an effective moment 
$\mu_{\text{eff}} \approx 1.77\,\mu_B$, corresponding to a Land\'e 
$g$-factor of $\approx\!2.04$, which is in good agreement with the 
expected value of $1.73\,\mu_B$ for a spin-$1/2$ system ($g \approx 2$).
These $\theta_d$ compare well with 
reported values of isotropic $\theta$ 
from Curie-Weiss fittings of susceptibility 
from experimental single crystal and powder, and theoretical data.
These $\theta$
fall within the range of $-70.1<\theta<-13.1$ $K$
\cite{pasco2019tunable,liu2024possible,gao2023discovery,haraguchi2017magnetic,haraguchi2024molecular,sun2022observation,sheckelton2017rearrangement,zhou2025antiferromagnetic,mangeri2024magnetoelectric}, 
with the exception of the reported ferromagnetic Weiss temperatures
of $+15$ $K$ (from single crystals) \cite{kim2023terahertz} and $+31$ $K$ (from DFT) \cite{jiang2017exploration}.

Since all our $\theta_d<0$, we have short-range AFM correlations.
The system's clear AFM characteristics above the critical temperature suggests that it could be frustrated due to the triangular lattice structure, as an AFM configuration cannot be fully satisfied in such a geometry.
To quantify this, we calculate the frustration index $f$
\cite{ramirez1994strongly}:
\begin{equation}\label{eq:frustration_index}
f=-\frac{\theta_d}{T_N}.
\end{equation}
From our susceptibility plots (e.g., Fig. \ref{fig:sus_x}), we found
$T_N\sim\! 20.2\pm 0.5$ $K$ (same for all $x,y,z$ directions within the range of error, as expected). Then, 
we have $(f_x,f_y,f_z)\sim\!(2.3,2.3,2.4)\pm0.1$, which 
indicates frustration since $f>1$.
This frustration may be a stepping stone towards
realizing exotic magnetic phenomena, such as quantum spin liquids.

We note that the calculated DMI vectors are consistent with the symmetry constraints imposed by the crystal (Moriya’s rules) and, upon enforcing the full little group at the bond midpoint, reduce to $D^1_x = 0$ for the NN bond in direction $(-1,0,0)$, $D^2_z = 0$ for the 2NN bond in direction $(-1,1,0)$, and $D^3_x = 0$ for the 3NN bond in direction $(-2,0,0)$, where the bond directions are expressed in the $(a,b,c)$ lattice basis (see Fig. \ref{fig:lattice}). 
See Supplementary Material C for a visualization of the DMI.
The origin of the DMI symmetry constraints can be understood by considering the site symmetry (little group) at each bond midpoint and applying Moriya’s rules for the DMI. For the NN and 3NN bonds oriented along the $a$-axis, the relevant vertical mirror plane in P3m1 is perpendicular to the bond and passes through its midpoint. This operation exchanges the two end sites, and the invariance of the antisymmetric exchange term $\mathbf{D}_{ij}\!\cdot(\mathbf{S}_i\times\mathbf{S}_j)$ requires the DMI vector to be invariant under this mirror. Because $\mathbf{D}$ transforms as an axial vector, the component along the bond direction changes sign under reflection and must vanish, hence $D^1_x=D^3_x=0$. In contrast, for the 2NN bond along $(-1,1,0)$, the bond midpoint lies on both the perpendicular-bisector mirror (which exchanges the sites) and a mirror that contains the bond (which does not).
The first again enforces $\mathbf{D}\!\perp$ bond; the second requires $\mathbf{D}$ to be odd under reflection, i.e. parallel to the in-plane normal of that mirror (perpendicular to the bond and within the plane), which eliminates the out-of-plane component: $D^2_z=0$.

Remarkably, the DMI
is comparable in magnitude
to the components of the anisotropic exchange:
The ratio between the Frobenius norm 
of $J^1$ ($|J^1|=2.92$) and Euclidean norm of $\mathbf{D^1}$ ($|\mathbf{D^1}|=0.90$) is $|\mathbf{D^1}|/|J^1|=0.31$. 
This implies the strength of the DMI is
relatively large, as the typical value of this 
ratio is on the order of $0.1$ \cite{xiang2013magnetic,moriya1960anisotropic}. 
We see that the role 
of the DMI may be significant, implying consequences such as the 
stabilization of 
non-collinear spin structures.
\\

\section{Strain tuning}
Given the potential for frustration in Nb$_3$Cl$_8$, we explore its magnetism further using biaxial strain tuning to better understand and manipulate its magnetic characteristics.
We expect the S=1/2 trimerization to persist under biaxial strain since the symmetry is not broken.
We used the four-state energy mapping technique
to calculate magnetic anisotropy parameters
as we applied compressive and tensile strain.
We applied biaxial strain in our DFT calculations
by changing the $a$ and $b$ lattice vectors
by the same percentage, and relaxing atom positions before running self-consistent field calculations for four-state mapping.
For computational efficiency in generating exchange vs. strain diagrams, we use a slightly lower 
energy cutoff for the plane-wave basis set (350 eV)
compared to the cases that we explore using Monte Carlo
calculations (400 eV) in Section \ref{sec:frustrationANDdmi}. We see from Supplementary Material
D that the exchange parameters extracted in both cases are similar to each other.

\begin{figure}[htbp]
    \centering
    \begin{subfigure}[t]{0.45\textwidth}
        \centering
        \includegraphics[width=\textwidth]{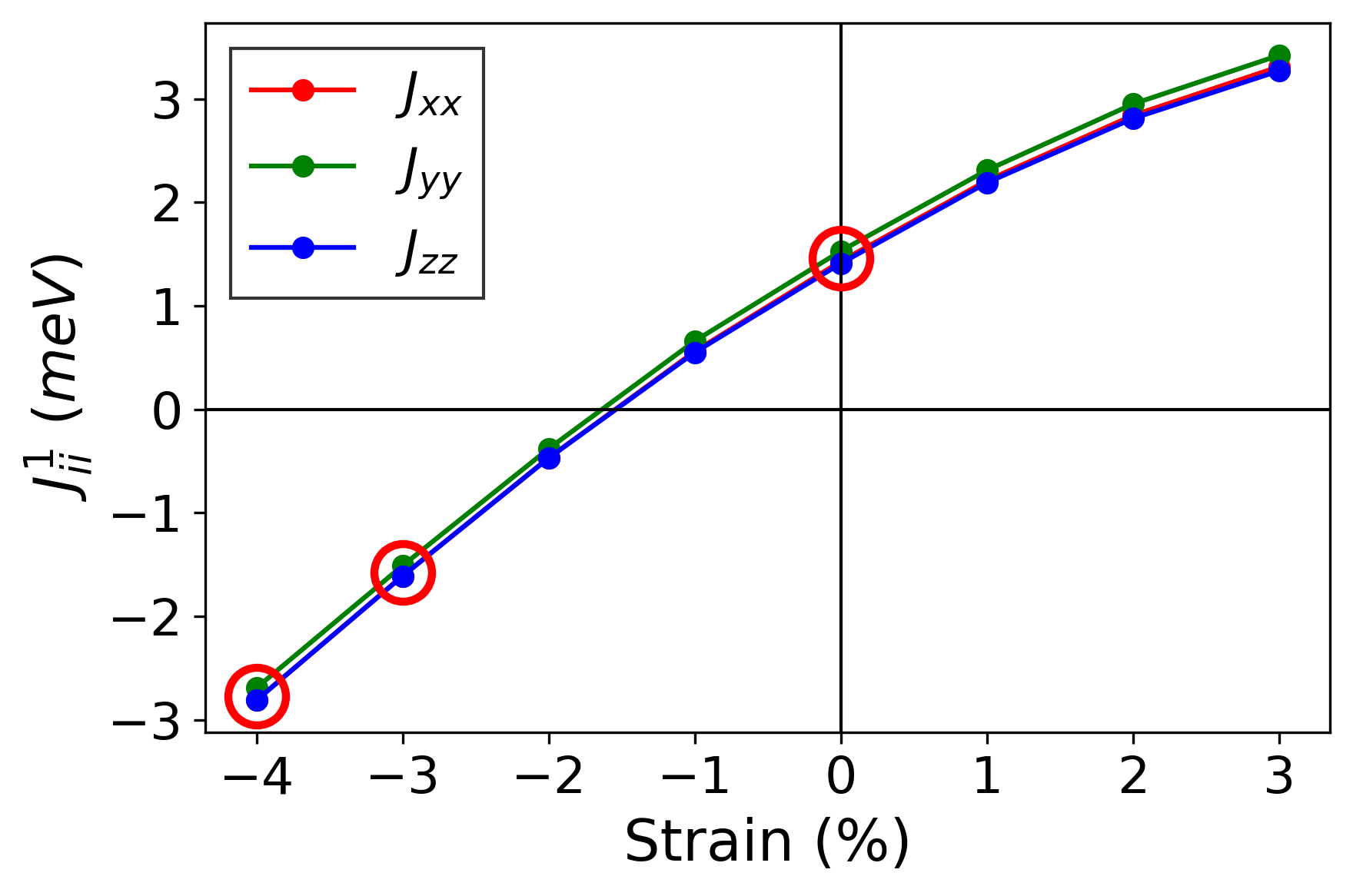}
        \caption{Diagonal matrix elements of $J^1$ (meV) vs strain ($\%$)}
        \label{subfig:diagJ1}
    \end{subfigure}
    \hspace{0.05\textwidth}
    \begin{subfigure}[t]{0.45\textwidth}
        \centering
        \includegraphics[width=\textwidth]{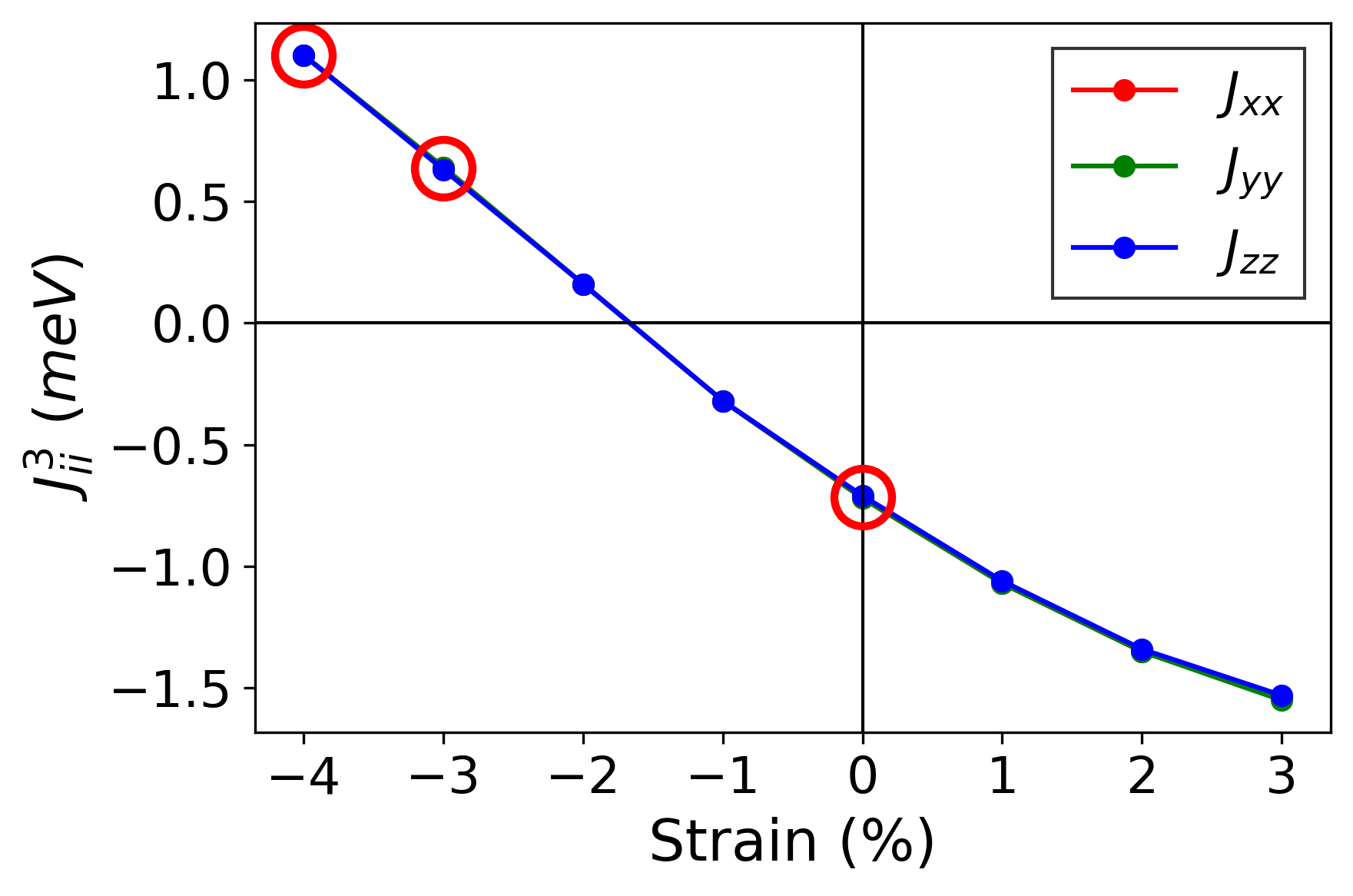}
        \caption{Diagonal matrix elements of $J^3$ (meV) vs strain ($\%$)}
        \label{subfig:diagJ3}
    \end{subfigure}
    \caption{Diagonal matrix elements $J_{xx},J_{yy}$ and
    $J_{zz}$ of $J^1$ and $J^3$, with respect to strain 
    (horizontal axis). 
    Note that some of the $J_{xx},J_{yy},J_{zz}$ curves in each case heavily overlap with each other.
    The red circles highlight the three test cases 
    explored in our Monte Carlo calculations: $0\%$, $-3\%$ and  $-4\%$.}
    \label{fig:diagComparison}
\end{figure}

We find that
the diagonal components of $J^1$ and $J^3$ in Eq. \eqref{eq:results_J123} 
can change sign with compressive strain,
indicating switching between AFM and FM interactions, as illustrated in Fig. \ref{fig:diagComparison}.
Other magnetic interaction parameters, including components of the DMI vector, can change sign and magnitude due to strain as well. We discuss those results in the Supplementary 
Material E.

Due to the change of magnetic exchange, the biaxial strain tunes 
the short-range magnetic correlations
from antiferromagnetic, paramagnetic to ferromagnetic, 
evident from the magnetic susceptibility data within three test cases:
$0\%,-3\%$ and $-4\%$. For $0\%$ strain, we have
the results in Eq. \eqref{eq:results_J123}, and for 
$-3\%$, we got single-ion anisotropy $A=0.46$ meV and,
using the same convention in Eq. \eqref{eq:J-matrix-elems},
the anisotropic exchange:
\begin{equation}\label{eq:results_J123_updated_m3}
\begin{array}{c}
\begin{aligned}
J^1 = 
\begin{bmatrix}
 -1.57 & -0.34 &  0.61 \\
  0.33 & -1.47 &  0.00 \\
 -0.62 &  0.00 & -1.57
\end{bmatrix},
J^2 = 
\begin{bmatrix}
  0.71 & 0.00 &  0.06 \\
  0.00 & 0.72 & -0.03 \\
 -0.07 & 0.04 &  0.71
\end{bmatrix}
\end{aligned}
\vspace{5pt}
\\[1ex]
\begin{aligned}
J^3 &= 
\begin{bmatrix}
  0.63 & 0.00 & 0.00 \\
  0.00 & 0.63 & 0.00 \\
  0.00 & 0.00 & 0.63
\end{bmatrix}
\end{aligned}
\end{array}
\end{equation}
For $-4\%$ strain, we got single-ion anisotropy 
$A=0.43$ meV and anisotropic exchange:
\begin{equation}\label{eq:results_J123_updated_m4}
\begin{array}{c}
\begin{aligned}
J^1 = 
\begin{bmatrix}
  -2.79 & -0.38 &  0.48 \\
   0.39 & -2.69 &  0.01 \\
  -0.47 &  0.01 & -2.79
\end{bmatrix}, 
J^2 = 
\begin{bmatrix}
   0.63 &  0.00 &  0.06 \\
   0.00 &  0.62 & -0.04 \\
  -0.06 &  0.03 &  0.62
\end{bmatrix}
\end{aligned}
\vspace{5pt}
\\[1ex]
\begin{aligned}
J^3 &= 
\begin{bmatrix}
   1.10 & 0.00 & 0.00 \\
   0.00 & 1.10 & 0.00 \\
   0.00 & 0.00 & 1.10
\end{bmatrix}
\end{aligned}
\end{array}
\end{equation}

Using these parameters in Monte Carlo simulations,
for $-3\%$ strain, we got 
$(\theta_x,\theta_y,\theta_z)=$ $(-0.6 \pm 0.4,$ $0.8 \pm 0.4,$ $-1.2 \pm 0.4)$;
and for $-4\%$ we got $(12.6 \pm 0.4,$ $12.4 \pm 0.4,$ $11.6 \pm 0.4)$.
See Fig. \ref{fig:sus_x} and Fig. S4 for the Monte Carlo susceptibility and fitting data. 
Since $\theta_d$ for $-3\%$ strain are very close to 0, it is reasonable to infer that
we are close to paramagnetism. 
For $-4\%$ strain, we clearly have at least short-range FM correlations since
$\theta_d>0$.
The phonon dispersions in Supplementary Material G show that the structures are stable.

The above results on strain-dependent exchange interactions and short-range correlations do not themselves establish the presence or absence of long-range magnetic order in the ground state.
To further assess the magnetic tendencies, we directly compared the total energies of representative magnetic configurations.

We perform \emph{ab initio} supercell and spin-spiral calculations 
as consistency checks
to show that the unstrained and $-4\%$ strained structures have 
energies close to the energies of standard commensurate
120$^\circ$ AFM and FM states, respectively. 
For the unstrained structure, this is consistent with recent works
which suggest that monolayer Nb$_3$Cl$_8$ has a 120$^\circ$ AFM ground state \cite{grytsiuk2024nb3cl8,aretz2025strong,mangeri2024magnetoelectric},
where the 120$^\circ$ order is defined on the triangular lattice of Nb$_3$ trimers (each treated as an effective $S={1}/{2}$ moment), i.e., the relative spin angles are between \emph{intertrimer} Nb$_3$ units rather than between \emph{intratrimer} Nb atoms.
However, due to the significant DMI we observe, it is 
reasonable to expect the true ground states to be incommensurate.

For the \emph{ab initio} supercell calculations, we choose
possible magnetic configurations that are commensurate with the supercell cell: FM, stripe (using a $2\times 2$ supercell),
and 120$^\circ$ order ($\sqrt3\times\sqrt3$ supercell). 
With SOC included, we tested several different spatial and orientational alignment of spins in each configuration (e.g., 120$^\circ$ with spins pointing inwards vs. outwards for a reference triangle; stripes along horizontal and vertical directions; and in-plane vs out-of-plane FM configurations).
We found that
for $0\%$ strain, the 120$^\circ$ configurations yielded the lowest energy per unit cell, with stripe configurations being higher by about $0.39$ meV per unit cell, and the FM configurations being higher than 120$^\circ$ order by about 1.55 meV.
In comparison, for $-4\%$ strain, the FM configurations have the lowest energy, with the 120$^\circ$ configurations being higher by about 2.05 meV per unit cell, and stripe configurations being higher than FM by about 2.45 meV.
Thus, while genuinely incommensurate states are inaccessible in single-supercell calculations, the energetics of the accessible commensurate configurations point to ground-state order close to 120$^\circ$ AFM at 0\% strain and FM at $-4$\% strain.

\begin{figure*}
    \centering
    \includegraphics[width=0.48\textwidth]{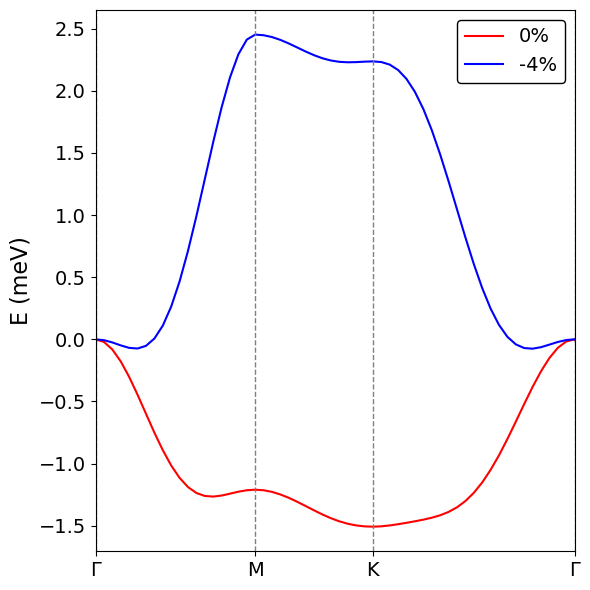}
    \caption{    
        Spin-spiral energies as a function of ordering vector for $0\%$
        and $-4\%$ biaxial strain, with the energy at $\Gamma$ set to $0$.}
    \label{fig:SS}
\end{figure*}

Next, we perform additional calculations based on spin-spirals without SOC, following the approach of Ref.~\cite{mangeri2024magnetoelectric} (see Supplementary Material H for more information). As seen in Fig.~\ref{fig:SS}, for $0\%$ strain, the minimum of the spin-spiral dispersion occurs at the $K$ point, which is consistent with 120$^{\circ}$ AFM order. For $-4\%$ strain, the minimum lies close to the $\Gamma$ point (0.08~meV away from the energy at $\Gamma$, with the total energy variation between the maximum and minimum being 2.52~meV).
This implies a tendency towards FM order. These trends agree qualitatively with the strain-tunable magnetic behavior demonstrated by the four-state mapping and supercell calculations with SOC. 

All these observations suggest that Nb$_3$Cl$_8$ exhibits a strain-tunable magnetic response, making it a promising material to explore short-range AFM, PM, and FM correlations for future applications.

\section{Conclusion}
Our findings provide deeper insights into the magnetic behavior of Nb$_3$Cl$_8$.
We showed that monolayer Nb$_3$Cl$_8$ may 
be magnetically frustrated. 
This result supports recent work that 
suggests exotic magnetic phenomena in 
Nb$_3$Cl$_8$.
Even though our classical Monte Carlo calculations 
do not capture quantum fluctuations, the AFM interaction and  frustration index $f>1$
indicate the possibility of Nb$_3$Cl$_8$ being a  
quantum spin liquid,
due to the spin-1/2 nature of the system 
and the strong 
quantum fluctuations.
These ground states 
are very exciting due to the wide 
array of potential applications in exploring 
fundamental quantum mechanics, quantum computing,
spintronics, and the development of advanced materials with novel magnetic and electronic properties.
We further showed that the short-range 
correlations
may be tuned between AFM, PM and FM
using strain, 
which is appealing due to the demand for
controllable magnetic properties in quantum materials 
technologies.
A potential next step would be to identify the role of the observed 
relatively-significant DMI,
and confirm whether it can stabilize exotic non-collinear 
spin structures such as spin spirals or skyrmions.
Further explorations could also involve
establishing a comprehensive phase diagram for the system.
By elucidating the role of magnetic anisotropy and extended interactions, our work paves the way for further theoretical and experimental studies aimed at harnessing the unique properties of Nb$_3$Cl$_8$ for applications in condensed matter research, device engineering, and beyond.

\section{Acknowledgments}
{This material is primarily based upon work supported by the National Science Foundation under Award No. DMR-2339995. Part of the work used advanced algorithm and codes supported by the Computational Materials Sciences Program funded by the U.S. Department of Energy, Office of Science, Basic Energy Sciences, Materials Sciences, and Engineering Division, PNNL FWP 83557. This research used resources of the National Energy Research Scientific Computing Center, a DOE Office of Science User Facility supported by the Office of Science of the U.S. Department of Energy under Contract No. DE-AC02-05CH11231 using NERSC award BES-ERCAP0033256. This work was facilitated through the use of advanced computational, storage, and networking infrastructure provided by the Hyak supercomputer system and funded by the University of Washington Molecular Engineering Materials Center at the University of Washington (NSF MRSEC DMR-2308979).}

\bibliography{references}

\newpage
\section*{Supplementary Material}

\appendix

\section{Density Functional Theory calculation details}\label{app:DFT}
An \emph{ab initio} density functional theory (DFT) approach was implemented to perform first-principles DFT+U+SOC calculations
using the \emph{Vienna Ab initio Simulation Package} (VASP) version 6.4.1 \cite{kresse1996efficiency,kresse1996software,kresse1993ab,kresse1994hafner,kresse1994ab},
for Hubbard $U$ and spin-orbit coupling (SOC).
The valence electron and core interactions were described using the projector augmented wave (PAW) method \cite{blochl1994projector,kresse1999ultrasoft}.
For the exchange-correlation energy functional, we used the local density approximation (LDA)
\cite{kohn1965self}. 
Within the applied PAW pseudopotentials,
we included semicore $4s$ and $4p$ states as valence
electrons for Nb,
and only the standard $3s$ and $3p$ states as valence electrons for Cl. 
We used a $4\times 4\times 1$ supercell
for the monolayer \footnote{We caution the reader that since we used a $4\times 4\times 1$ supercell, we are not excluding 
interactions between
periodic images when calculating 
magnetic anisotropy parameters between third-nearest-neighbors (i.e., $J^3_{ij}$). 
So, we deviate from the standard four-state energy mapping technique in the 3NN case by 
accounting for this change
in coordination number 
by dividing 
all computed $J^3_{ij}$ by $2$ since 
each interaction occurs twice for this choice of supercell.
A $5\times 5\times 1$ (or larger) supercell would have mitigated 
this issue. However, we could only converge our calculations for the collinear Heisenberg case with a $5\times 5\times 1$ supercell 
despite experimenting with a wide range
of \emph{LAMBDA} and Wigner-Seitz radii for the noncollinear case. 
Despite this, we confirm that the isotropic Heisenberg $J^4$ from 4NN interactions is negligible using the collinear
case. This justifies truncating interactions
past 3NN interactions even using the four-state mapping method. This is consistent 
with other works like Ref. \cite{mangeri2024magnetoelectric} considering interactions only until 3NN, and our own 
spin-spiral calculations mapping to the spin-1/2 Heisenberg model for the case without SOC (which gave $J^4=0$ meV, as seen in Supplementary Material \ref{app:SS}).}.
Since lattice parameter information from experiment
is not available for the monolayer, we used the parameters 
$a=6.744$ $\text{\AA}$ and $\alpha=120^{\circ}$
from single-crystal X-ray
diffraction data for $\alpha$-Nb$_3$Cl$_8$ \cite{magonov1993scanning},
which is the multi-layer high-temperature structure
that is proposed to lead to exotic spin phenomena
and remain in the $\alpha$ phase at low temperatures
if exfoliated at high temperatures
\cite{grytsiuk2024nb3cl8,liu2024possible,pasco2019tunable}.
A vacuum spacing of $25 \text{\AA}$ along the $c$-direction was applied to avoid interactions between 
periodic images of the monolayers.
We used a $1\times 1\times 1$ $\Gamma$-centered 
mesh with automatically-determined $k$ points.
The conjugate-gradient algorithm was used 
to relax only the atom positions until the norms 
of all the forces were smaller than
$10^{-7}$ eV$/\dot{\text{\AA}}$. 
To account for the in-plane anisotropy, all Nb magnetic moments were constrained along the $x$ direction during structural relaxation. Prior to computing the single-ion anisotropy constants, we confirmed the in-plane anisotropy by comparing total energies for several collinear configurations in which all Nb moments were constrained to point along the same direction $\hat{\mathbf{m}}=(m_x,m_y,m_z)$.
The resulting energies increase in the following order:
$E_{(1,0,0)}$ $=$ $E_{(0,1,0)}$ $=$ $E_{(-1,1,0)}$ $=$ $E_{(1,1,0)}$ $=$ $E_{(-\sqrt{3},-1,0)}$ $<$ $E_{(1,-\sqrt{3},0)}$ $=$ $E_{(-1/2,\sqrt{3}/2,0)}$ $<$ $E_{(0,1,1)}$ $=$ $E_{(1,0,1)}$ $<$ $E_{(0,0,1)}$.
Taking $E_{(1,0,0)}$ as the reference, the energy differences $\Delta E = E - E_{(1,0,0)}$ (in meV) for the above directions are:
$0$, $0$, $0$, $0$, $0$, $0.020$, $0.021$, $0.050$, $0.050$, $0.081$.
Because the relaxed crystal structure is insensitive to the choice among the lowest-energy initial magnetic-moment configurations, we fixed all Nb moments along the $x$ direction during structural relaxation as a convenient and representative choice.

We set the magnetic moments of Nb$_3$ trimers by 
assigning to each Nb atom the same initial magnetic moment, to correspond to the desired moment of a triangle vertex point.
In order to accurately capture the shared
spin-$1/2$ moment, we constrained the direction and sign of the assigned moment for each spin configuration, but not the size
(which we let the software determine).
The convergence criterion of the total energy was
$10^{-8}$ eV.
The energy cutoff for the plane-wave basis set 
was $400$ eV (except for the $J_{ij}^k$ vs. strain plots 
Fig. 3 (of the main text), and Fig. \ref{fig:diagComparisonSupp},\ref{fig:diagComparisonSupp2NN} here, which use $350$ eV; see main text for clarification).
We used Gaussian smearing, with a smearing width of
$0.03$ eV.
The Wigner-Seitz radii for Nb and Cl
were $1.270$ $\text{\AA}$ and $1.111$ $\text{\AA}$
respectively; and the \emph{LAMBDA} parameter (which
sets the weight with which the penalty terms of the constrained local moment approach enter into the total energy expression and the Hamiltonian) was set to $9$. 
The simplified (rotationally invariant) approach to DFT+U introduced by Dudarev et al. \cite{dudarev1998electron}
was used to incorporate the DFT+U formalism.
A Hubbard potential of $U = 1$ eV was used on Niobium d orbitals to account for effective
on-site Coulomb interactions. 
Our optimized structure yielded an Nb-Nb bond length 
of $2.8112$ $\text{\AA}$ within an Nb$_3$ trimer. This is only 
$0.01\%$ different from the experimental bond length of
$2.8109$ in Ref. \cite{haraguchi2024molecular,haraguchi2017magnetic},
although it is quite different from the $2.85$ $\text{\AA}$
bond length in Ref. \cite{xiong2025role}.

\section{Monte Carlo calculation details}\label{app:MC}

For our classical Monte Carlo calculations, we used VAMPIRE 6.0 \cite{evans2014atomistic} with
the parameters computed using DFT. 
Due to the system's anisotropy, we do not
expect $\chi_d$ to always be the same for different $d$.
$\theta_d$ can be easily found by fitting the linear 
portion of $\chi_d^{-1}$ vs. $T$ data to a line, and 
calculating $T$ at which $\chi_d^{-1}=0$. This is simply 
the negative of the vertical intercept $b$ divided by the 
gradient $m$: $-b/m$. 
The propagated uncertainty $\sigma_\theta$ of $\theta_d$ (without the small covariance term) is calculated using 
$
\sigma_\theta \approx \sqrt{\Bigl(\tfrac{\sigma_b}{m}\Bigr)^2 \;+\;\Bigl(\tfrac{b\,\sigma_m}{m^2}\Bigr)^2}
$, where $\sigma_b$ is the standard error of $b$,
and $\sigma_m$ is the standard error of the $m$.
$C$ in Eq. (2) (of the main text) is the inverse of the gradient, and then 
the effective magnetic moment is $\mu_{\text{eff}}=\sqrt{8C}\mu_B$
(for the Bohr magneton $\mu_B$). 
For a spin-$1/2$ moment, we expect $\mu_{\text{eff}}\approx 1.73\mu_B$
\cite{mugiraneza2022tutorial}.
When fitting our data to a line, we used the data only above $50$ $K$, in order to ensure we are  within the 
paramagnetic region for which the Curie-Weiss law is 
applicable, well above the kinks in susceptibility that arise at or below $\sim\!20$ $K$ in all our cases (e.g., in Fig. 2 (of the main text)).
In our susceptibility plots Fig. 2 (of the main text) and Fig. \ref{fig:susyz}, 
the equations of the linear fits are given in the inset
in the top-left along with the coefficient of determination $R^2$. The $R^2$ values very close to $1$ indicate 
that the fits explain the variance in  
data quite well. 

Unless otherwise stated,
we employed a 
2D triangular lattice of size $100 \times 100$ nm,
with periodicity in the x and y directions. 
This yields $25628$ atoms for the unstrained structure,
27234 atoms for the $-3\%$ biaxial strained structure,
and 27900 atoms for the $-4\%$ biaxial strained structure.
We modeled the triangular lattice in VAMPIRE using a 2D rectangular `unit cell' of dimensions $a\times a\sqrt{3}\times 1000$ (for lattice vector $a$ and `vacuum spacing' $1000$),
with two Nb$_3$ `atoms' placed at $(0.25,0.75,0)$
and $(0.75,0.25,0)$ in fractional coordinates.
We used temperature increments of $0.2$ $K$, 
$2.5\times 10^6$ equilibration time steps, and
$2.5\times 10^6$ loop time steps. 
We additionally assigned the initial spins to be in 
random directions.

\section{Origin of DMI constraints}
\label{app:DMI_symmetry}

\begin{figure}[H]
    \centering
    \includegraphics[width=0.48\textwidth]{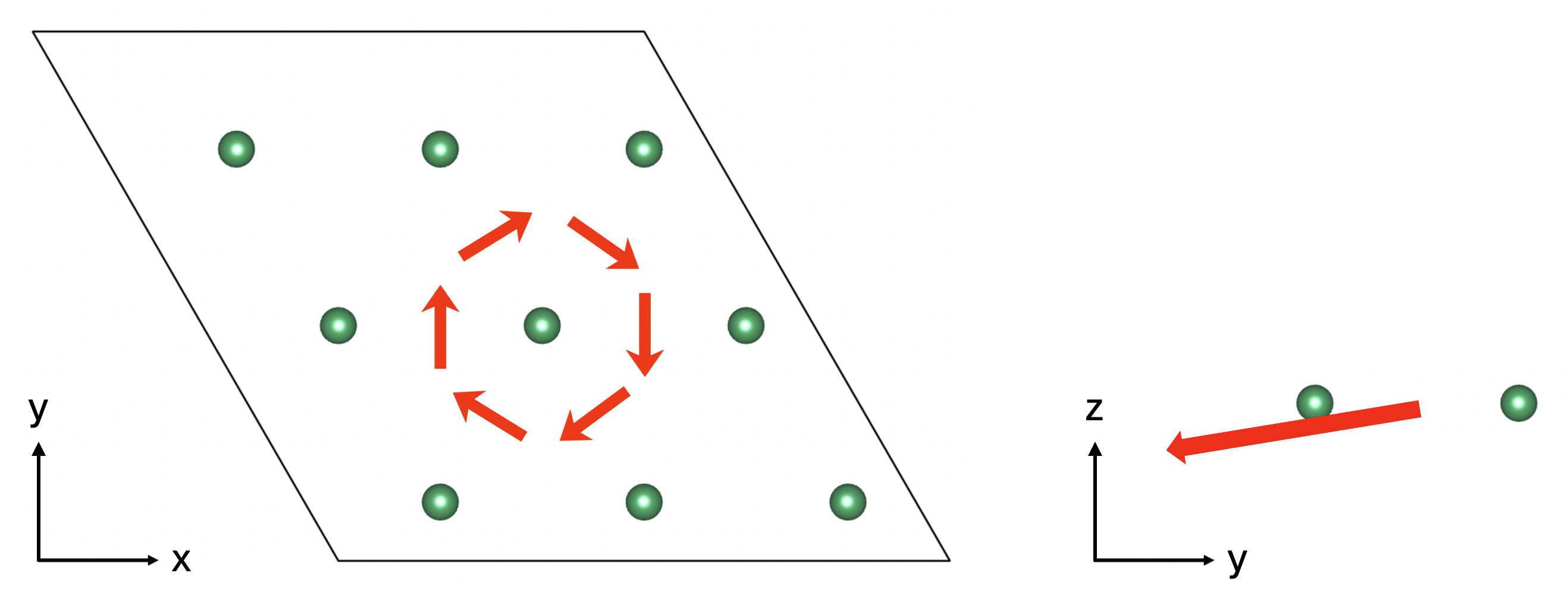}
    \caption{    
            $\mathbf{D}^1$ for the unstrained case, visualized in 
            the $xy$ (left) and $yz$ (right) planes. The $D_z$ component is $-0.15$ meV for all NN bonds. 
    }
    \label{fig:DMI:supp}
\end{figure}

The DMI vectors extracted from Eq. (3) (of the main text) 
for the unstrained case are
$\mathbf{D}^1=(0, -0.89, -0.15)$ meV,
$\mathbf{D}^2=(-0.02, -0.04, 0)$ meV,
and 
$\mathbf{D}^3=(0, 0, 0)$ meV. $\mathbf{D}^1$ is visualized in
Fig. \ref{fig:DMI:supp}.

\section{Anisotropic exchange and single-ion anisotropy for unstrained structure with plane-wave basis energy cutoff of 350 eV}\label{app:ENCUT350}

Using a plane-wave basis energy cutoff of 350 eV, we get a single-ion anisotropy of 0.56 meV (same as 
with 400 eV), and the anisotropic exchange in Eq. \eqref{eq:results_J123_350}. These values are very close to the 
values for $400$ eV given in the main text Eq. (3) (of the main text).

\begin{equation}\label{eq:results_J123_350}
\begin{array}{c}
\begin{aligned}
J^1 = 
\begin{bmatrix}
  1.44 & 0.16 & -0.88 \\
  -0.16 & 1.52 & 0.00 \\
  0.88 & 0.00 & 1.40
\end{bmatrix},
J^2 = 
\begin{bmatrix}
  0.96 & 0.00 & 0.04 \\
  0.00 & 0.96 & -0.04 \\
  -0.04 & 0.04 & 0.96
\end{bmatrix}
\end{aligned}
\vspace{5pt}
\\[1ex]
\begin{aligned}
J^3 &= 
\begin{bmatrix}
  -0.72 & 0.00 & 0.00 \\
  0.00 & -0.72 & 0.00 \\
  0.00 & 0.00 & -0.72
\end{bmatrix}
\end{aligned}
\end{array}
\end{equation}

\section{Strain tuning data}\label{app:strain}

\begin{figure*}[t!]
    \centering
    \begin{subfigure}[t]{0.45\textwidth}
        \centering
        \includegraphics[width=\textwidth]{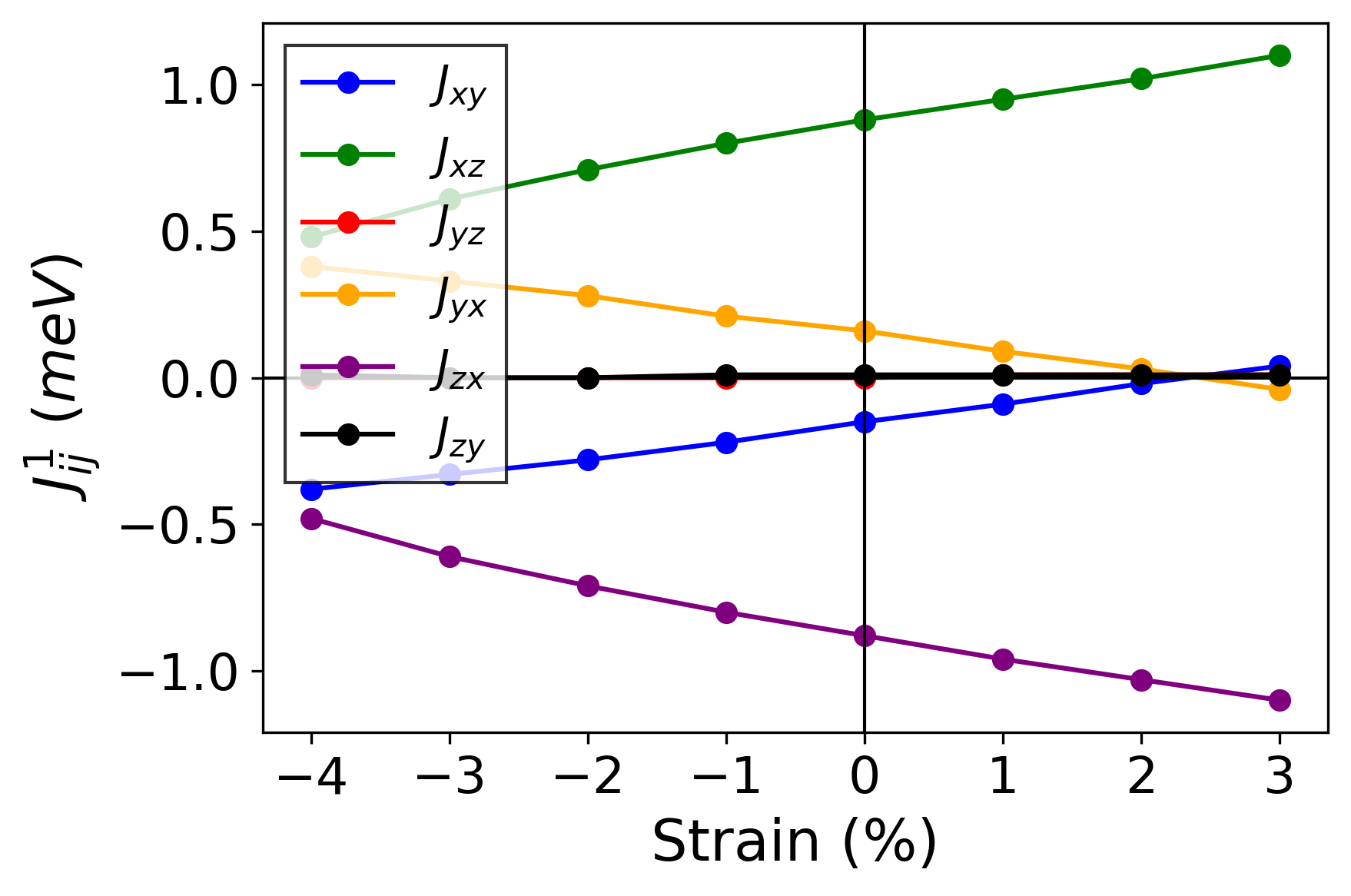}
        \caption{Non-diagonal matrix elements of $J^1$ (meV) vs strain ($\%$)}
        \label{subfig:nondiagJ1}
    \end{subfigure}
    \hspace{0.05\textwidth}
    \begin{subfigure}[t]{0.45\textwidth}
        \centering
        \includegraphics[width=\textwidth]{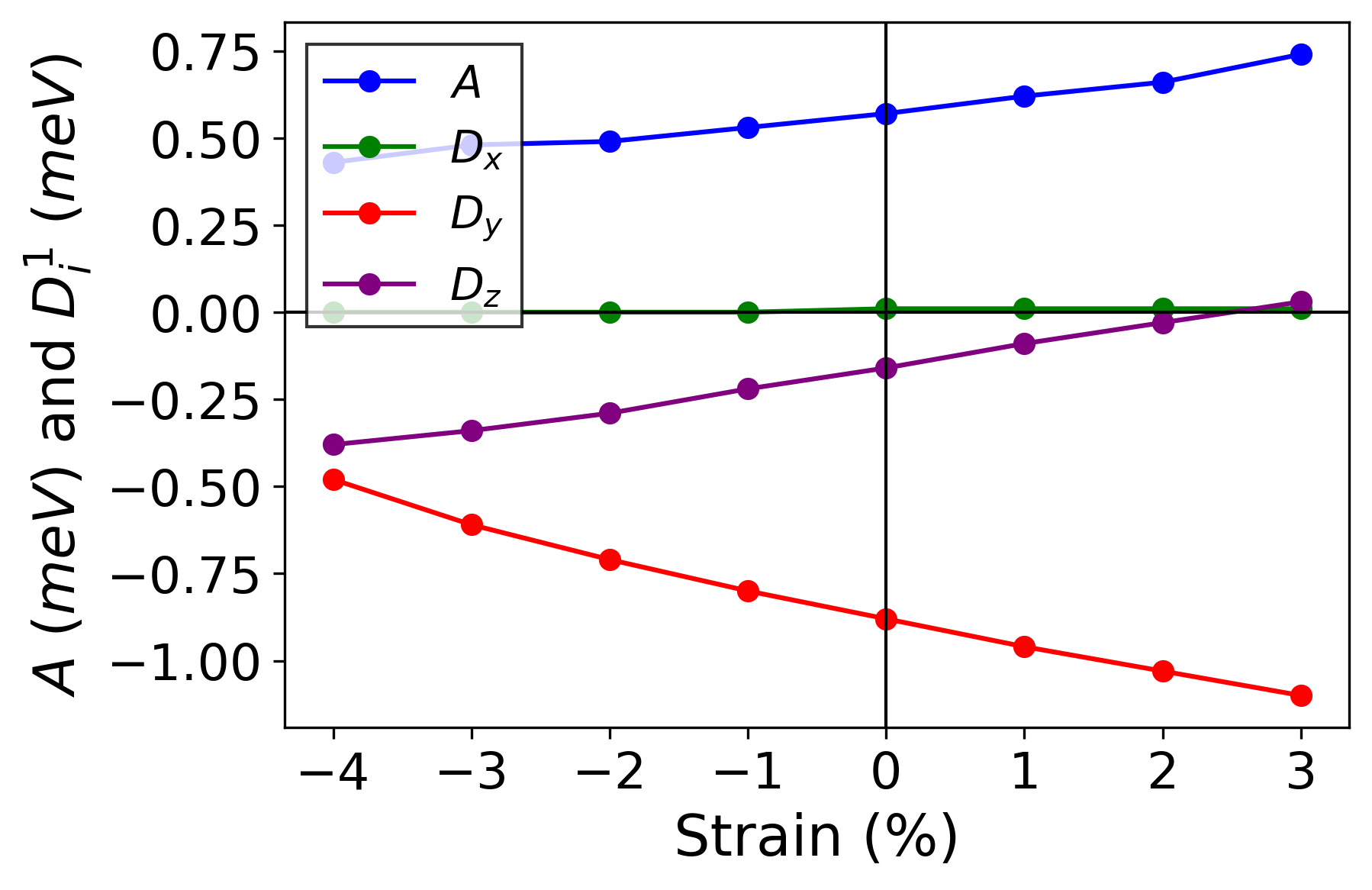}
        \caption{Components of DMI vector $\mathbf{D}^1=(D_x,D_y,D_z)$ (meV) and single-ion anisotropy $A$ (meV) vs strain ($\%$)}
        \label{subfig:DMIJ1}
    \end{subfigure}

    \caption{Non-diagonal elements of $J^1$, DMI from NN interactions, and single-ion anisotropy,
    with respect to strain. 
    }
    \label{fig:diagComparisonSupp}
\end{figure*}

\begin{figure*}[t!]
    \centering
    \begin{subfigure}[t]{0.45\textwidth}
        \centering
        \includegraphics[width=\textwidth]{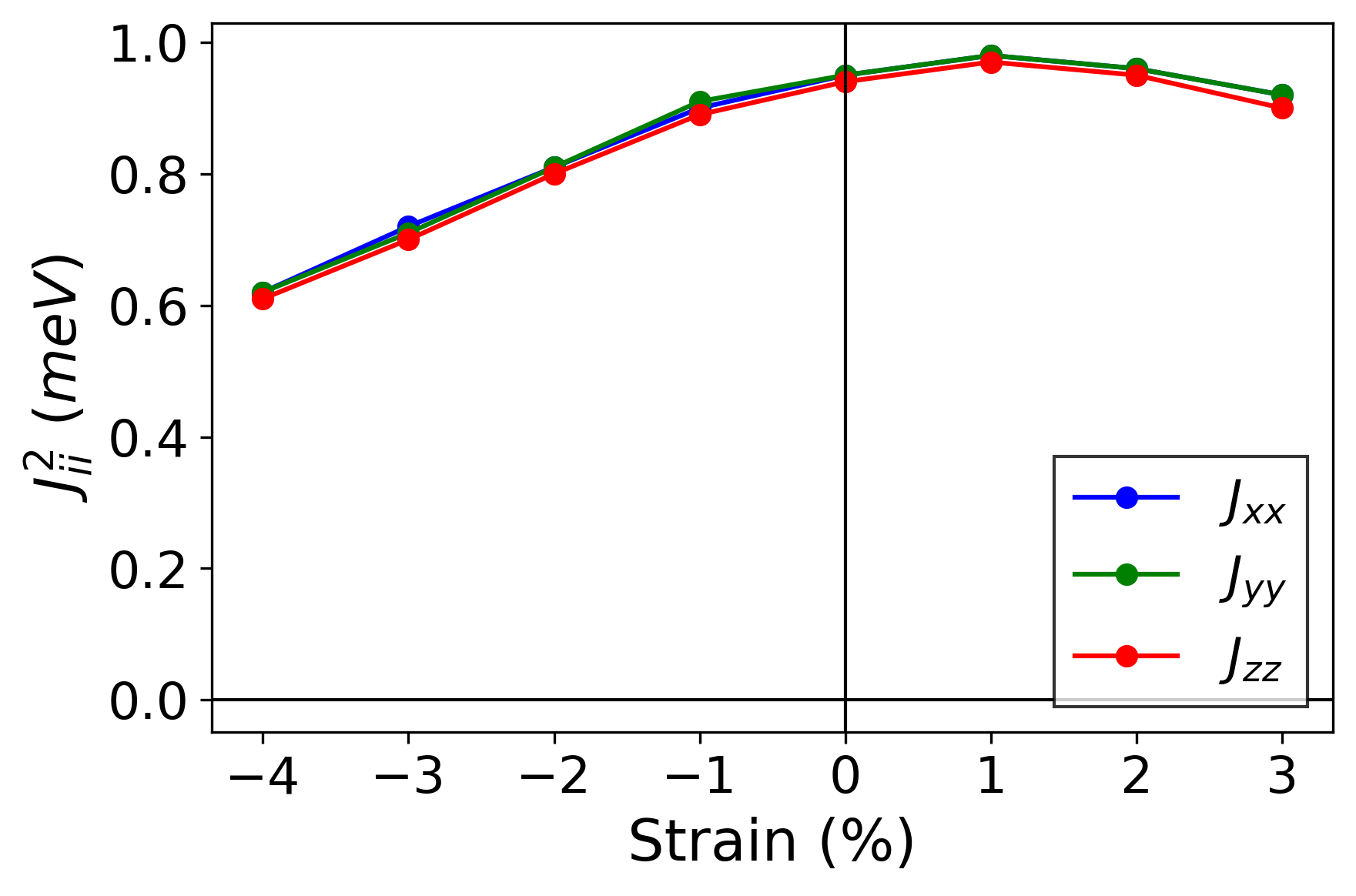}
        \caption{Diagonal matrix elements of $J^2$ (meV) vs strain ($\%$)}
        \label{subfig:diagJ2}
    \end{subfigure}
    \hspace{0.05\textwidth}
    \begin{subfigure}[t]{0.45\textwidth}
        \centering
        \includegraphics[width=\textwidth]{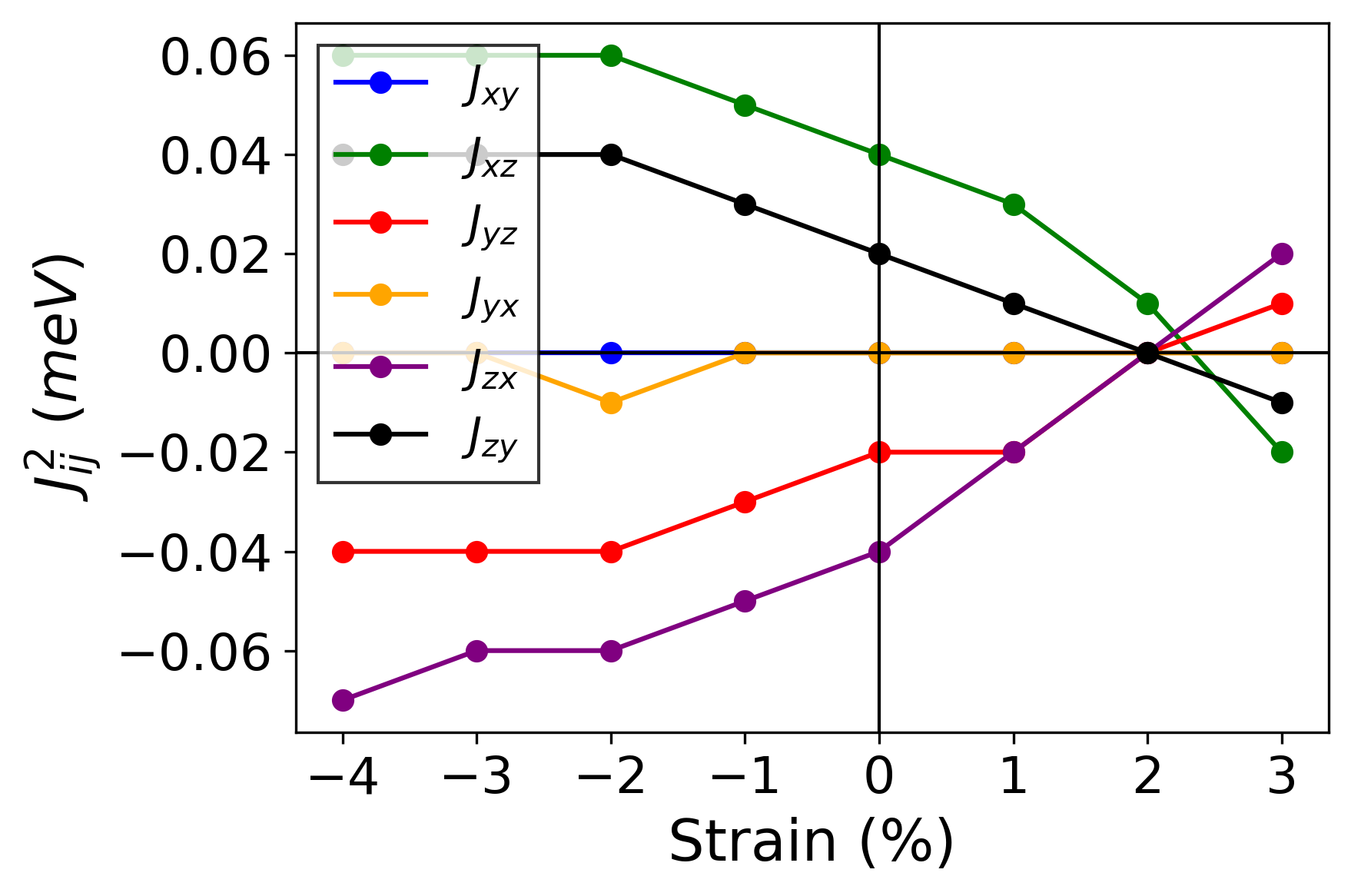}
        \caption{Non-diagonal matrix elements of $J^2$ (meV) vs strain ($\%$)}
        \label{subfig:nondiagJ2}
    \end{subfigure}
    \hspace{0.05\textwidth}
        \begin{subfigure}[t]{0.45\textwidth}
        \centering
        \includegraphics[width=\textwidth]{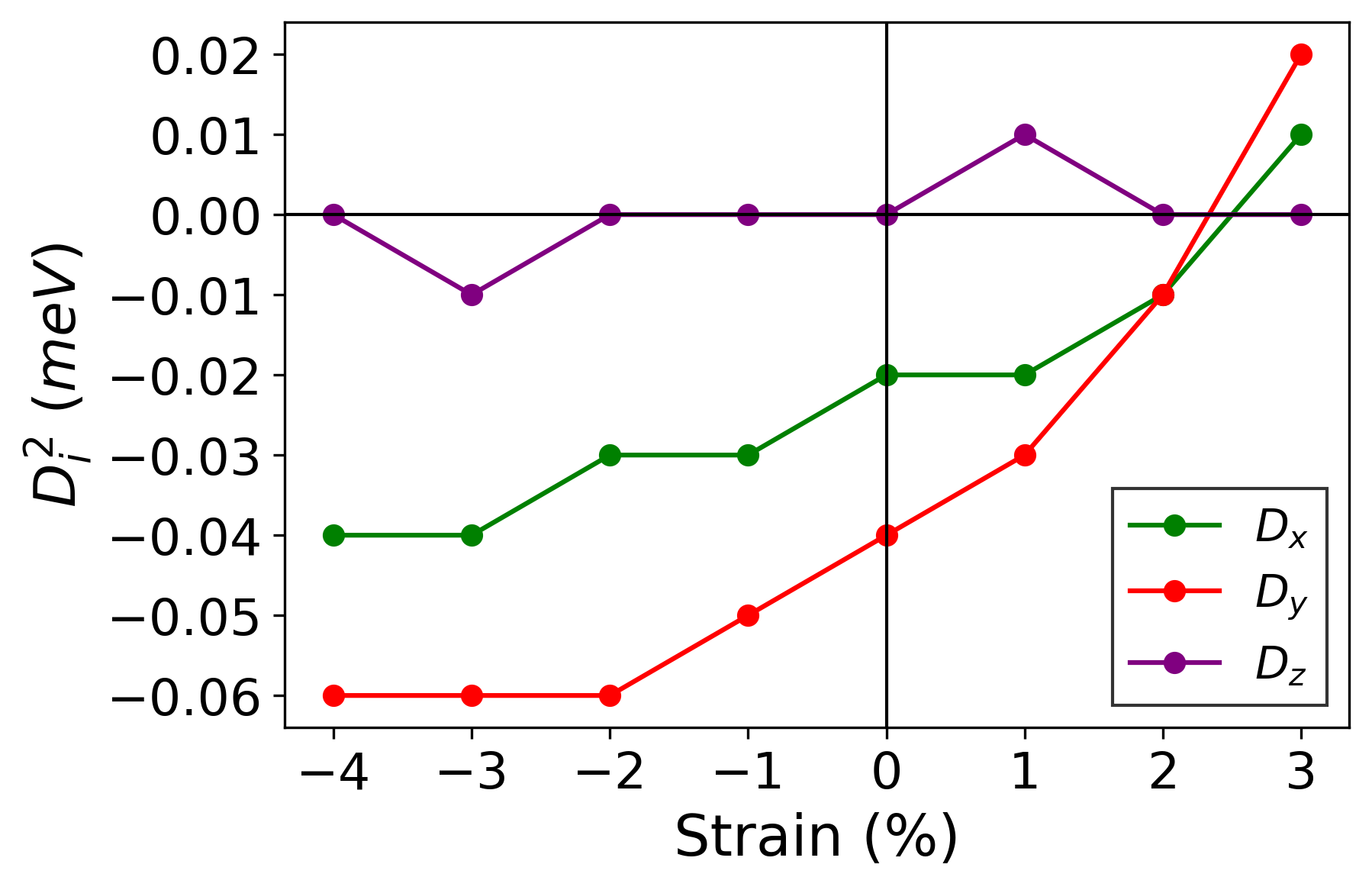}
        \caption{Components of DMI vector $\mathbf{D}^2=(D_x,D_y,D_z)$ (meV) vs strain ($\%$)}
        \label{subfig:DMIJ2}
    \end{subfigure}
    \caption{Diagonal elements of $J^2$, non-diagonal elements of $J^2$, and DMI from 2NN interactions, 
    with respect to strain. 
    }
    \label{fig:diagComparisonSupp2NN}
\end{figure*}

We present results on the behavior of
magnetic anisotropy parameters that are excluded
from the main text for NN interactions and single-ion anisotropy
in Fig. \ref{fig:diagComparisonSupp},
and for 2NN interactions in Fig. \ref{fig:diagComparisonSupp2NN}.
Note that for all percentages of strain applied, the non-diagonal components of $J^3$ 
(and therefore DMI) are $0$
up to our 2 decimal place accuracy.

\section{Additional susceptibility data}

\begin{figure*}[t!]
    \centering
    \begin{subfigure}[t]{0.48\linewidth}
        \centering
        \includegraphics[width=\linewidth]{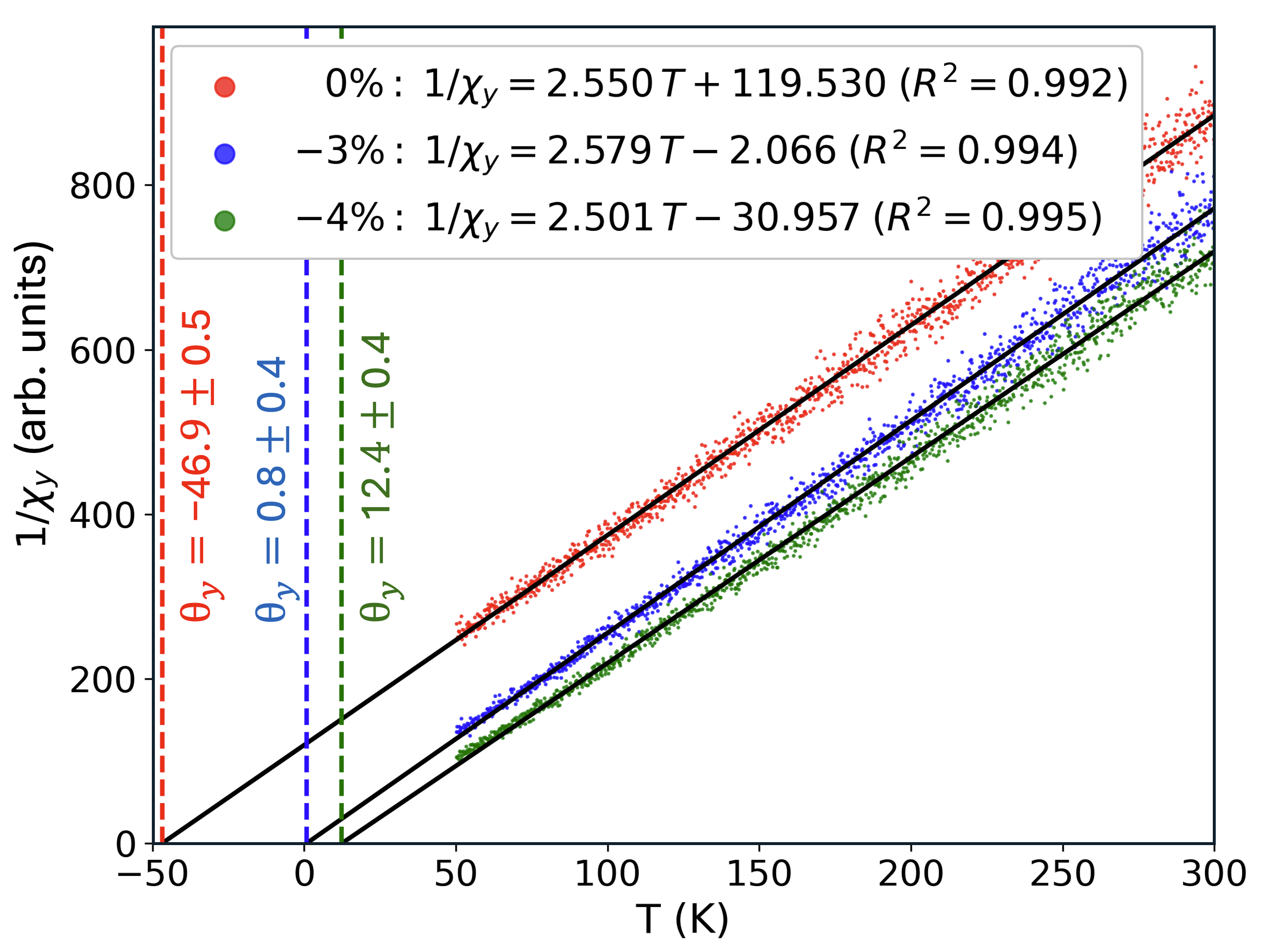}
        \caption{$d=y$}
        \label{subfig:susy}
    \end{subfigure}
    \begin{subfigure}[t]{0.48\linewidth}
        \centering
        \includegraphics[width=\linewidth]{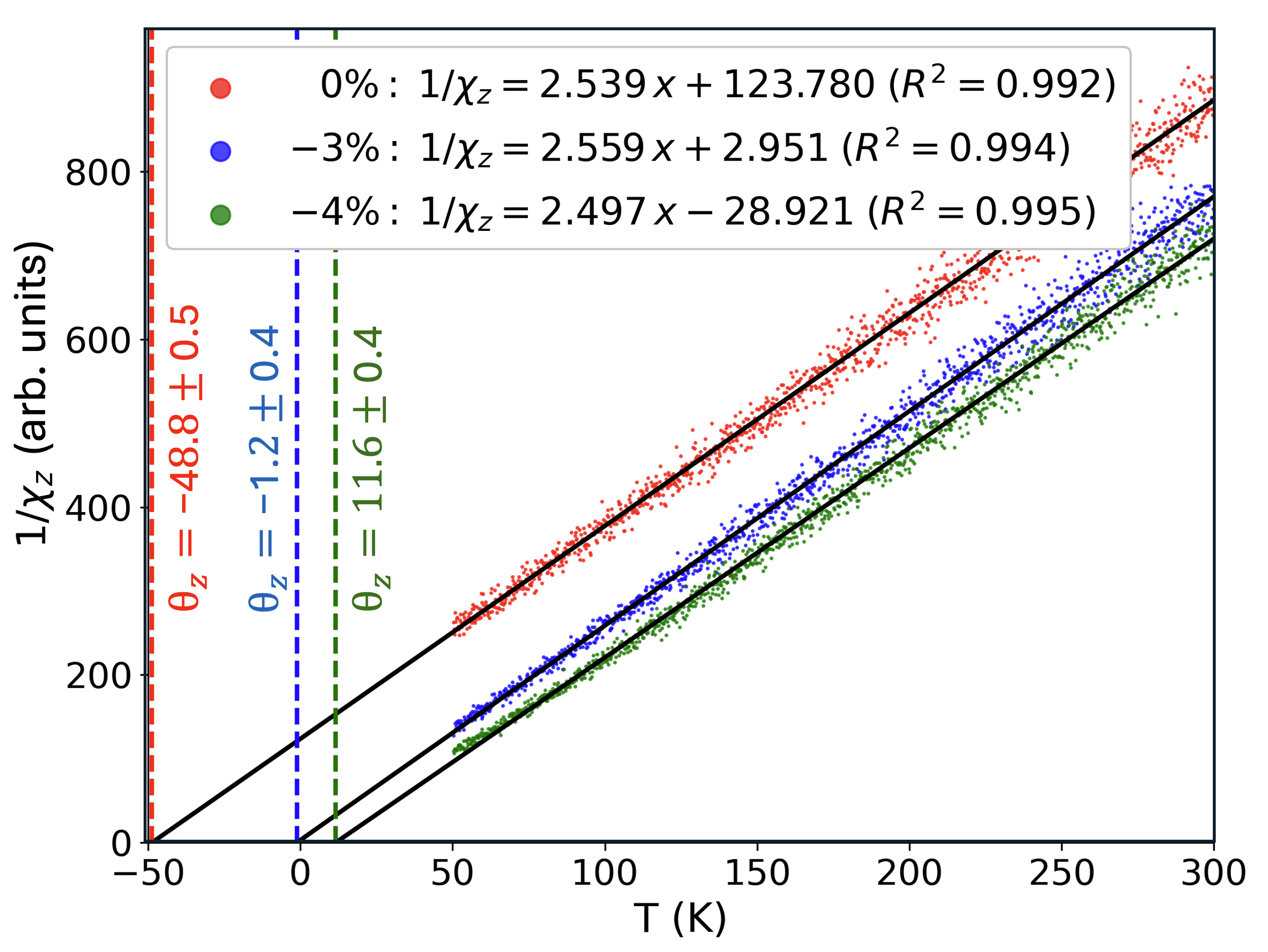}
        \caption{$d=z$}
        \label{subfig:susz}
    \end{subfigure}
    \caption{
    Inverse susceptibility $1/\chi_d$ (arb. units) vs. temperature $T (K)$ in the $d=y$ and $d=z$-directions for
        $0\%$, $-3\%,$ and $-4\%$ biaxial strain.  
        Notation is as used in Fig. 2 (of the main text).
    }
    \label{fig:susyz}
\end{figure*}

We present susceptibility data for the 
y and z-directions
in
Fig. \ref{fig:susyz}.

\section{Phonon dispersions showing structural stability}\label{app:phonons}

We use phonon dispersions to show that our structures are stable for the $0,-3\%$ and $-4\%$ biaxial strained cases.
We used VASP to perform Density Functional Perturbation Theory (DFPT) calculations for the monolayer unit cell with SOC, with unconstrained Nb magnetic moments starting in the $(1,0,0)$ direction. Otherwise, the calculation details are as in Supplementary Material \ref{app:DFT}.

\begin{figure*}[t!]
    \centering
    \begin{subfigure}[t]{0.45\textwidth}
        \centering
        \includegraphics[width=\textwidth]{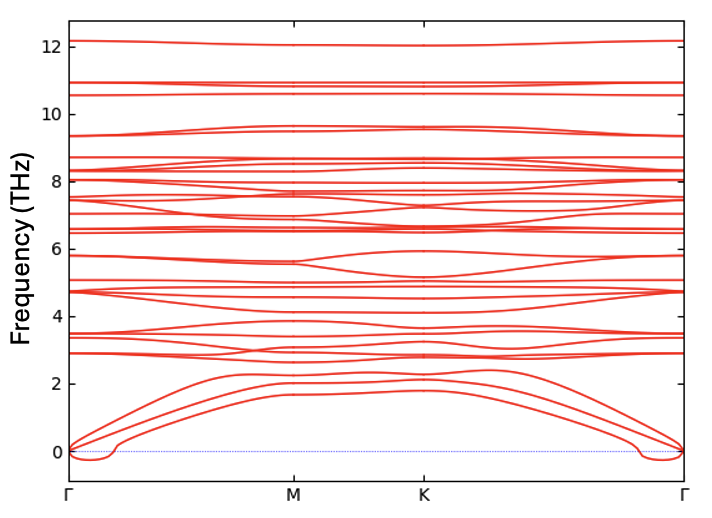}
        \caption{Unstrained structure}
        \label{subfig:phstr0}
    \end{subfigure}
    \hspace{0.05\textwidth}
    \begin{subfigure}[t]{0.45\textwidth}
        \centering
        \includegraphics[width=\textwidth]{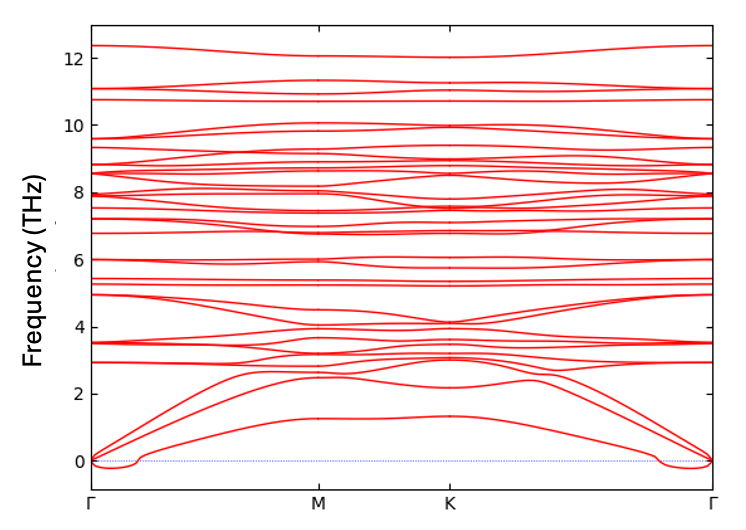}
        \caption{$-3\%$ biaxial strained structure}
        \label{subfig:phstrbim3}
    \end{subfigure}
    \hspace{0.05\textwidth}
    \begin{subfigure}[t]{0.45\textwidth}
        \centering
        \includegraphics[width=\textwidth]{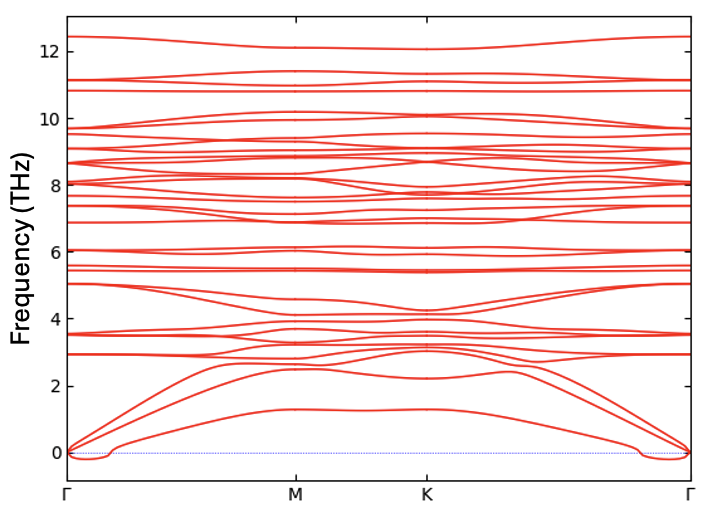}
        \caption{$-4\%$ biaxial strained structure}
        \label{subfig:phstrbim4}
    \end{subfigure}
    \caption{Phonon dispersions. 
    }
    \label{fig:diagPhonons}
\end{figure*}

Indeed, we see from Fig. \ref{fig:diagPhonons} that the phonon frequencies are positive and imply stable structures. Note that there is a small pocket of negative frequencies near $\Gamma$, with 
the minima of the negative frequencies being 
$\sim-0.27$ THz for $0\%$,
$\sim-0.22$ THz for $-3\%$ and 
$\sim-0.21$ THz for $-4\%$.
These small values lie within the $\sim-1$ THz  threshold commonly attributed to numerical noise near the $\Gamma$ point \cite{lin2022general,libbi2020thermomechanical}, and is therefore not indicative of a true structural instability.

\section{Spin-spiral calculations}\label{app:SS}

We performed spin-spiral calculations without SOC, because VASP's spin-spiral implementation uses the generalized Bloch theorem: Adding SOC couples spin and lattice degrees of freedom, and thereby breaks the spin-rotation invariance required for the spiral boundary conditions. Due to the lack of SOC, 
the results in this section are intended only as a qualitative reference and not as a replacement for the SOC-based results in the main text.

We computed the spin-spiral dispersion for the unstrained
and $-4\%$ biaxial strained structures. The spiral energies were fitted to an isotropic spin-1/2 $J_{1}–J_{2}–J_{3}–J_{4}$ Heisenberg model of the following form (using notation consistent with Eq. (1) (of the main text)):
\[
H = \sum_{i<j} J^{k}_{ij}\, \mathbf{S}_{i}\cdot\mathbf{S}_{j},
\]
from which excellent linear fits were obtained (see Fig. 4 (of the main text)). For $0\%$ strain the best fit parameters were
\[
(J^{1}, J^{2}, J^{3}, J^{4}) = (0.99,\; 0.22,\; 0.35,\; 0.00)\ \text{meV},
\]
and for $-4\%$ strain the parameters were
\[
(J^{1}, J^{2}, J^{3}, J^{4}) = (-2.72,\; 0.27,\; 0.73,\; 0.00)\ \text{meV}.
\]

The extracted exchange constants differ from those obtained from the four-state mapping method because SOC is not included in the spin-spiral calculations and therefore the anisotropic contributions to the exchange constants are absent. However, in both cases, we still see dominant AFM $J^1$ for $0\%$ strain, and dominant FM $J^1$ for $-4\%$ strain.

\section{Band diagrams and magnetism}\label{app:bands}

\begin{figure*}[htbp]
    \centering
    \begin{subfigure}[t]{0.45\textwidth}
        \centering
        \includegraphics[width=\textwidth]{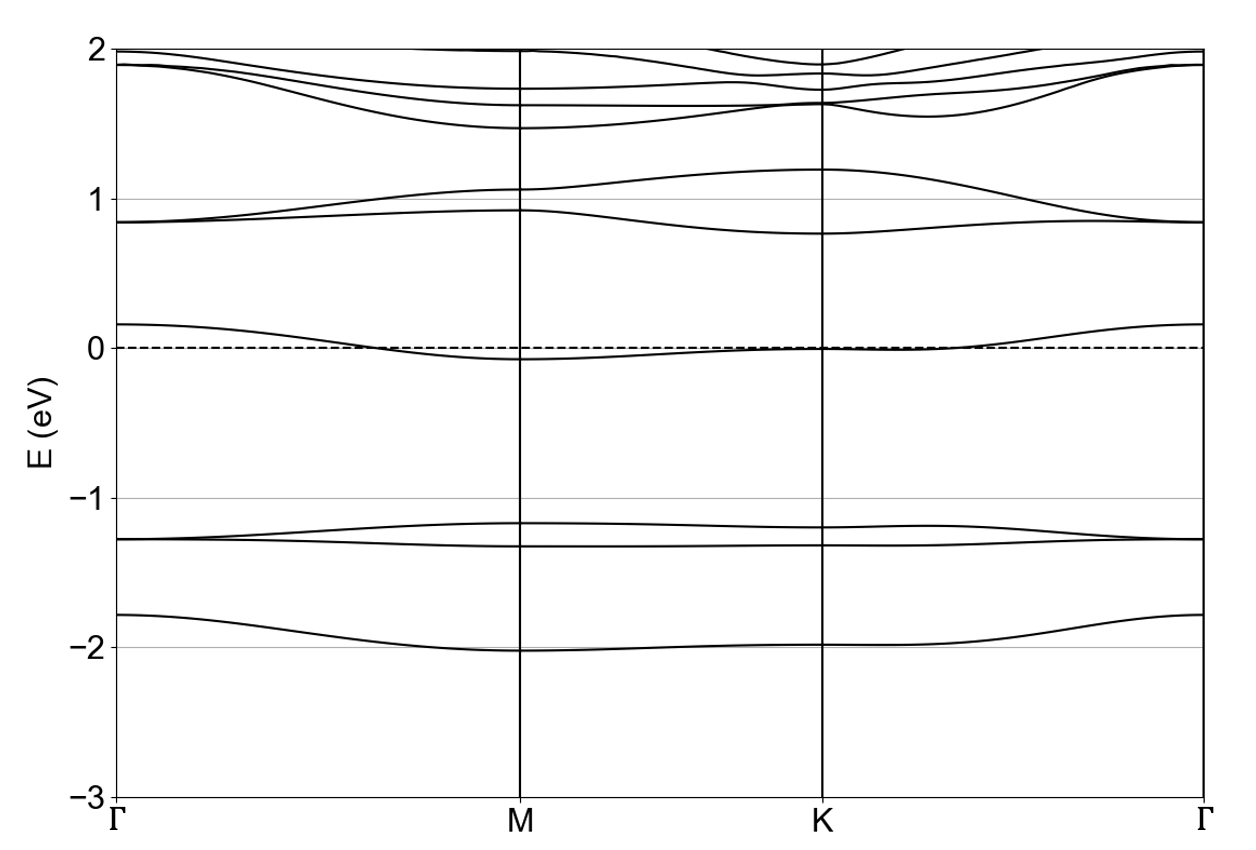}
        \caption{No spin polarization.}
        \label{subfig:bands_nonmag}
    \end{subfigure}
    \hspace{0.05\textwidth}
    \begin{subfigure}[t]{0.45\textwidth}
        \centering
        \includegraphics[width=\textwidth]{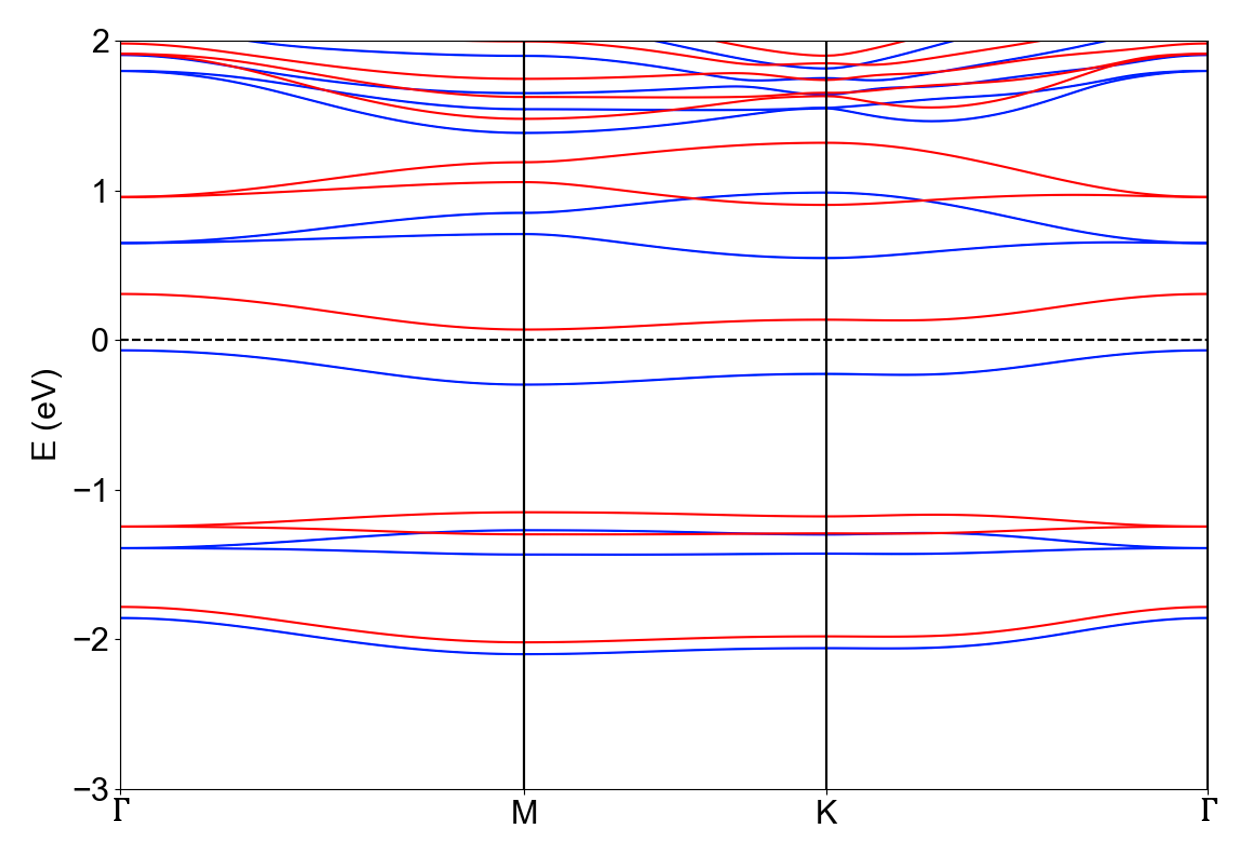}
        \caption{Spin polarized, but without SOC.
        Blue bands are the spin up channel, and red bands spin down.}
        \label{subfig:bands_spin}
    \end{subfigure}
    \hspace{0.05\textwidth}
    \begin{subfigure}[t]{0.45\textwidth}
        \centering
        \includegraphics[width=\textwidth]{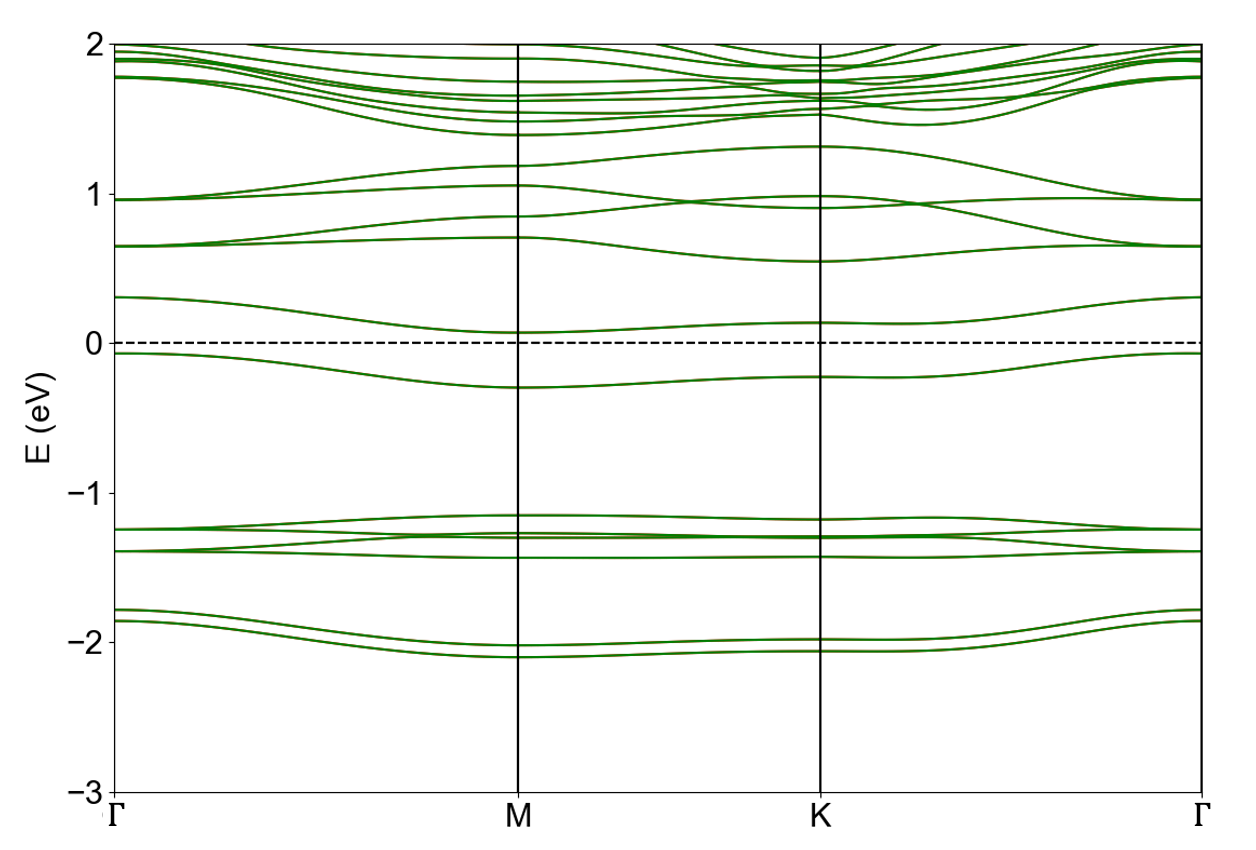}
        \caption{With SOC. Bands are doubly-degenerate.}
        \label{subfig:bands_SOC}
    \end{subfigure}
    \hspace{0.05\textwidth}
    \begin{subfigure}[t]{0.45\textwidth}
        \centering
        \includegraphics[width=\textwidth]{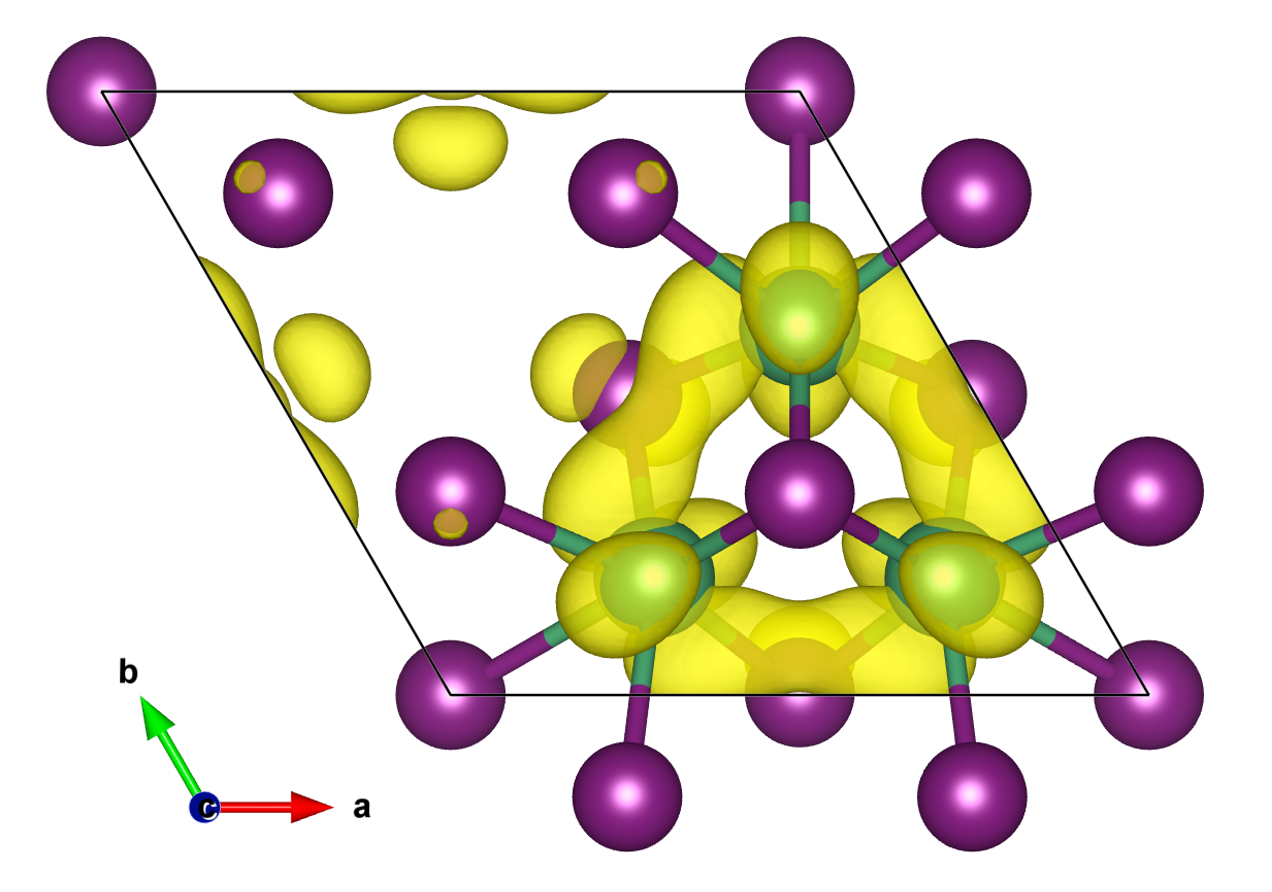}
        \caption{Magnetization in monolayer Nb$_3$Cl$_8$.}
        \label{subfig:mag}
    \end{subfigure}
    \caption{(a)-(c) Band diagrams for monolayer Nb$_3$Cl$_8$ along the path $\Gamma-M-K-\Gamma$ in the Brillouin zone.
    (d) Magnetization (yellow) in monolayer Nb$_3$Cl$_8$ unit cell, viewed from above the $ab$-plane.
        The magnetization is calculated by subtracting the 
        spin up density by the spin down density.
        Nb atoms are in green, and Cl atoms in purple.
        Most of the magnetization is from the Nb atoms, 
        while the small yellow bubbles in the top-left are 
        minor contributions from Cl atoms.
    }
    \label{fig:bands}
\end{figure*}

For completeness, we present band diagrams 
for monolayer Nb$_3$Cl$_8$ in
Fig. \ref{fig:bands} (a)-(c).
We note that our band diagram for the case 
without spin polarization Fig. \ref{fig:bands} (a)
compares well with 
diagrams from other references, such as 
Ref. \cite{grytsiuk2024nb3cl8,sun2022observation}
even though they used PBE as the exchange-correlation
energy functional (compared to the LDA we used).

We also present a plot of magnetization in Fig. \ref{fig:bands} (d).

\end{document}